\documentclass[12pt, draftclsnofoot, onecolumn]{IEEEtran}

\usepackage{preamble}

\newcounter{MYtempeqncnt}

\doublefalse
\extendedfalse
\arxivtrue
\redfalse
\bluefalse
\greenfalse

\begin{document}

\title{Throughput Optimization in FDD MU-MISO Wireless Powered Communication Networks}

\author{Arman~Ahmadian
\ifarxiv
\thanks{This work has been submitted to the IEEE for possible publication. Copyright may be transferred without notice, after which this version may no longer be accessible.}
\fi
}

\maketitle

\begin{abstract}

In this paper, we consider a frequency-division duplexing (FDD) multiple-user multiple-input-single-output (MU-MISO) wireless-powered communication network (WPCN) consisting of one hybrid data-and-energy access point (HAP) with multiple antennas which coordinates energy/information transfer to/from several single-antenna wireless devices (WD).
Typically, in such a system, wireless energy transfer (WET) requires such techniques as \emph{energy beamforming} (EB) for efficient transfer of energy to the WDs.
Yet, efficient EB can only be accomplished if channel state information (CSI) is available to the transmitter, which, in FDD systems is only achieved through uplink (UL) feedback.
Therefore, while in our scheme we use the downlink (DL) channels for WET only, the UL channel frames are split into two phases:
the CSI feedback phase during which the WDs feed CSI back to the HAP and the WIT phase where the HAP performs wireless information transmission (WIT) via space-division-multiple-access (SDMA).
To ensure rate fairness among the WDs, this paper maximizes the minimum WIT data rate among the WDs.
Using an iterative solution, the original optimization problem can be relaxed into two sub-problems whose convexity conditions are derived.
Finally, the behavior of this system when the number of HAP antennas increases is analyzed.
Simulation results verify the truthfulness of our analysis.

\end{abstract}

\begin{IEEEkeywords}
CSI feedback, WPCN, energy beamforming, throughput, rate fairness, doubly near-far effect, MU-MISO
\end{IEEEkeywords}

\IEEEpeerreviewmaketitle
\section{Introduction}
\ext{\subsection{Motivation}}
\IEEEPARstart{T}{he} limited lifetime of battery-powered wireless devices (WD)s in a conventional wireless communication network has always been a fundamental bottleneck for which the only solution has been to manually replace or recharge the batteries after depletion.
Yet, there exists applications where doing so is laborious or even impractical\footnote{Consider, for example, a large number of WDs deployed in a large area, or a sensor implanted in the human body.}.
Wireless-powered communication (WPC) has recently emerged as a promising solution to prolong the lifetime of such energy-constrained WDs \cite{07081084, 07009979}.
WPC can be used in a variety of applications, such as IOT networks, RFID systems, and wireless sensor networks (WSNs) \cite{07462480, 07462479}.
It uses RF-enabled wireless energy transfer (WET) \cite{07081084} as a means to wirelessly supply the energy of WDs, thus enabling them to function seamlessly without the need of battery replacement/ recharging.
WET can be defined as the use of electromagnetic (EM) waves to transfer energy from an energy transmitter (ET) to an energy receiver (ER) over the air \cite{07009979}. 

As with many new technologies, WET raises its own issues however.
First, since the power signal from the transmitter is severely attenuated over distance, the problem of transferring sufficient energy over even moderately large distances is not trivial.
Second, in many cases there are multiple ERs which could all be mobile.
Therefore, the scheme needs to be adaptable to multiple receivers and at the same time robust to mobile scenarios \cite{07081084}.
When multiple antennas are available at the ET, the solution is to carefully weight the transmitted signals from different antennas in such a way that they superimpose constructively at the ERs and destructively everywhere else \cite{07462480}.
This technique, referred to as \emph{energy beamforming} (EB), results in the concentration of the transmitted wave into narrow beams, and thus enables the ET to deliver ample energy to the ERs \cite{07009979, 06133872, 07511170}.

%

WET has been studied in a number of works.
In \cite{07393604}, the probability of outage, and in \cite{07401119, TSP.2016.2641400, 14930604},\comm{also 15201087} channel acquisition and training methods for WET systems have been studied.
In \cite{07511170} fairness-aware EB methods were investigated and in \cite{07009979} a pareto optimal energy beamformer (EB) was derived.
In what follows, we concentrate on a crucial application of WET, namely, WPC.

\subsection{Wireless Powered Communication}
WPC is the result of using WET in a wireless communication network whose purpose is wireless information transmission (WIT).
There are currently two different lines of research in WPC: simultaneous wireless information and power transfer (SWIPT) \cite{06133872} \comm{and 04595260} and wireless powered communication network (WPCN) \cite{07009979, 06678102}\comm{also 06954434}.
In SWIPT, both energy and information are carried via the same RF signal, while in WPCN the AP transfers energy to the WDs in downlink (DL), and the WDs perform uplink (UL) WIT using the harvested energy.
\arx{This means that, conceptually in SWIPT, data is transmitted \emph{to} the ER, whereas it is  transmitted \emph{by} the ER in case of WPCN.
Although in both cases of WPCN and SWIPT, the harvested energy decays rapidly with respect to the distance between the wireless device (WD) and the AP, power constraints for WPCN are more stringent than those of SWIPT.
This is so because, while in SWIPT the harvested energy is only needed to keep the WDs alive for energy harvesting (EH) and information decoding (ID), in WPCN the transmitted power in the UL WIT is supplied by the DL WET.
This is fundamentally more difficult as the WD is now required to actively transmit, rather than silently ``listen''.}{}

WPCNs have been recently studied in the literature.
References \cite{07081084} and \cite{07462480}, for instance, provide overviews of possible WPCN configurations and the techniques employed in such networks.
In what follows we briefly discuss the challenges encountered in designing a WPCN with regard to our design paradigm and introduce some of the relevant papers that discuss such issues.

\subsubsection{Duplexing}
The most fundamental challenge in WPCN (and SWIPT) design is the so-called \emph{energy-information trade-off}.
This trade-off exists because energy and information both use the same communication resources, such as time and bandwidth (BW).
Since in WPCNs information and energy are transferred in different directions, this trade-off is achieved via duplexing techniques in such networks.

\begin{itemize}

\item When time-division duplexing (TDD) is used \cite{07009979, 06678102, 06623072, 07564600, 1511.01291v2, 07332956, 07814311, 07906591, 08357530, 1807.05543, 1807.05670}\comm{also 06954434}, the challenge is to determine the optimal time lengths during which the UL WIT and the DL WET occur while the advantage is that channel reciprocity may be taken advantage of for channel state informtion (CSI) acquisition at the transmitter when multiple antennas are available there \cite{04050100}.
Yet, due to the orthogonal time allocations of the DL and UL channels, WET can not occur continuously.
This is a restriction for energy transmitters having a peak transmit power, limiting the total amount of delivered energy.

\item Although frequency division duplexing (FDD) has not been thoroughly studied in WPCNs, it has been more successfully implemented in wireless communication networks in general \cite{04050100}.
When used in WPCNs \cite{16638567, 06568923, 1807.05670}, the total available BW should be optimally allocated to the DL and UL channels.
However, since channel receiprocity does not apply to such a system, the CSI must be fed back to the access point \cite{01715541, 04050100, 06568923, 16638567}.
Hence, the optimum design should specify how much CSI feedback is needed to achieve the best performance.
In this scheme, however, the energy may be continuously transmitted, posing less restriction to the peak transmit power.

\end{itemize}

\subsubsection{Multiple Access}
While single-device scenarios have been considered in  \cite{06568923, 16638567, 07393872, 07814311, 1807.05543}, WPCNs may also be used for multiple WDs
\cite{07009979, 07332956, 1511.01291v2, 07906591, 08357530}\comm{also 06954434, 07386616}.
When multiple devices are considered, multiple access techniques should be utilized for the UL channels.
In \cite{06623072, 07332956, 08357530} TDMA is used for which the challenge is to optimally partition the frame length into orthogonal time slots serving different WDs.
When the HAP is equipped with multiple antennas, space-division-multiple-access (SDMA) may be utilized \cite{07009979, 06678102, 07564600}\comm{06954434} 
which enjoys a higher spectrum efficiency than TDMA.
Finally, in \cite{1511.01291v2} non-orthogonal multiple access (NOMA) technique has been employed.
\arx{Note that due to the intrinsic broadcast property of wireless transmission, the transmitted energy in the DL can be harvested by all WDs and multiple access techniques are pointless for WIT.
Nevertheless, when the HAP is equipped with multiple antennas, EB may be utilized to generate multiple beams aimed at multiple WDs.
This may be viewed as a multiple access technique for energy where the purpose is to send desired amounts of energy to each WD.}{}
\subsubsection{Fairness}
WPCNs can have separate APs for the purpose of information reception and energy transmission for which case the APs are referred to as \emph{data access points} (DAPs) and \emph{energy access points} (EAPs) respectively \cite{07393872}.
On the other hand, the DAP and EAP may be collocated in which case it is called a \emph{hybrid access point} (HAP).
References \cite{06678102, 07009979, 06623072, 1511.01291v2, 07814311, 07906591}\comm{also 07386616, 06954434} consider a system with a HAP, while in \cite{07393872, 07332956, 06568923} both scenarios are considered.
When HAPs are emplyed, the WDs located far from the HAP, harvest less energy in each block compared to the closer WDs.
Furthermore, if they are to enjoye the same SINR and hence throughput, WDs located far from the HAP should consume more power when transmitting data in the UL compared to closer WDs \cite{07462480}.
This creates a severely unfair situation referred to as the \emph{doubly near-far} effect \cite{06678102} which substantially starves far WDs of rate if rate fairness is not considered.
To address the this issue, in \cite{06678102, 07009979} maximizing the \emph{minimum rate}, among the WDs is considered.

%
\subsection{This Paper}
In this paper, a FDD multiple-user multiple-input-single-output (MU-MISO) WPCN consisting of one HAP with multiple antennas and a set of distributed single-antenna WDs, as illustrated in Fig. \myref[fig.schematic], is studied.

\begin{figure}[t]
\begin{center}
\centering
\ifdouble
\includegraphics[trim=90 190 500 170,clip, width=\textwidth]{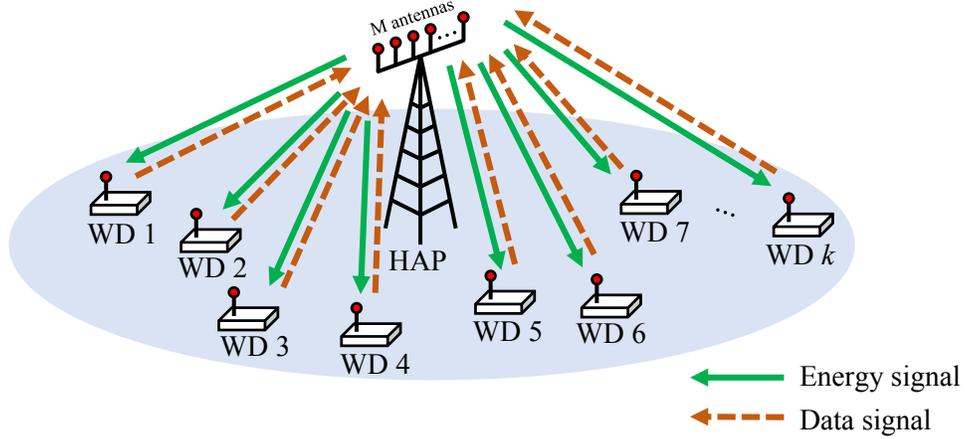}
\else
\includegraphics[trim=90 190 500 170,clip, scale=1]{schematic.eps}
\fi
\end{center}

\caption{\textblue{Schematic of a WPCN employing multiple antennas at the HAP.}}
\label{fig.schematic}
\end{figure}
%
We try to address all of the aforementioned challenges for a FDD MU-MISO WPCN; specifically

\begin{itemize}

\item Given a certain amount of available BW, the maximum allowed power spectrum density, and a fixed frame length, the optimal UL and DL BWs as well as the optimal amount of CSI feedback time lengths are calculated.

\item It is shown that, under finite-rate CSI feedback, the  beamformer for this problem can be pareto optimal when each WD sends at least one feedback bit.

\item To ensure rate fairness among the WDs, the optimization is performed such that the minimum throughput among the WDs, referred to as the minimum WIT data rate, is maximized.
In order to achieve this, the HAP allocates more wireless power to the WDs located farther away from the HAP relative to the others.

\item We will define a metric called the \emph{fairness radius} which describes a circular boundary around the HAP.
We will show that the WDs whose distances to the HAP are greater or equal to this radius achieve equal data rates and those whose distances are less than this radius attain higher data rates than the rest of the WDs.

\item It is shown that as the number of HAP antennas goes to infinity, the optimal CSI feedback phase time length ratio, the optimal DL BW ratio, and the fairness radius tend to zero.

\end{itemize}

\ifextended
\subsubsection{\color{olive}Notation}
\fi
\myitem{\ext{Uppercase and Lowercase:~}We will use bold lower-case and upper-case letters for column vectors and matrices respectively, non-bold lower or upper-case Latin or Greek letters for scalars, and caligraphic letters for sets.}
%
\myitem{\ext{Matrices and Vectors:~}For any arbitrary m by n matrix (m-vector) $\boldsymbol{A}(\boldsymbol{a})$, $\boldsymbol{A}^T(\boldsymbol{a}^T)$ and $\boldsymbol{A}^H(\boldsymbol{a}^H)$ represent its transpose and conjugate transpose respectively. \textblue{Defining sets $\mathcal{S} \subset \left\{1,...,m\right\}$, and $ \mathcal{T} \subset \left\{1,...,n\right\}$, submatrix $[\boldsymbol{A}]_{\mathcal{S},\mathcal{T}}$ (subvector $[\boldsymbol{a}]_{\mathcal{S}}$) is a matrix (vector) whose elements are $[\boldsymbol{A}]_{i,j}, \forall i\in \mathcal{S}, j \in \mathcal{T}$ ($[\boldsymbol{a}]_{i}, \forall i\in \mathcal{S}$) in the original column and row order (original order). }}
\myitem{\ext{Standard Vectors:~}$\boldsymbol{1}_n$, $\boldsymbol{0}_n$, and $\boldsymbol{e}_k$ will be used to represent all-one, all-zero and the $k$-$th$ canonical basis vector of $\mathbb{R}^{n}$
\textblue{(a vector of all zeros, except for the $k$-$th$ place, where it is one), while $\boldsymbol{I}_n$ will represent as an identity matrix of order $n$.}}
\myitem{\ext{Hadamard Product:~}For two vectors $\boldsymbol{a}$ and $\boldsymbol{b}$ of the same dimension, $\boldsymbol{a}\odot\boldsymbol{b}$ represents the element-wise or the Hadamard product.}
\myitem{\ext{Vector Operations:~}In a small abuse of notation, we interpret vector exponentiation in an element-wise fashion; i.e. for $\boldsymbol{a}\in \mathbb{C}^{K\times1}$ and $d\in \mathbb{C}$, vectors $\boldsymbol{b}=\boldsymbol{a}^d$ and $\boldsymbol{c}= d^{\boldsymbol{a}}$ are vectors of the same dimension as $\boldsymbol{a}$ for which $b_i=a_i^d,c_i=d^{a_i},~\forall i,~1\le i\le K$.}
\myitem{\ext{Positive Operator:~}Furthermore, for vector $\boldsymbol{a}\in\mathbb{R}^{K\times1}$, $\left\{\boldsymbol{a}\right\}^+$ is a vector of the same dimension for which $\{\boldsymbol{a}\}_k^+=\mathrm{max}\{a_k,0\}$ and for vector $\boldsymbol{a}\in \mathbb{C}^{K\times1}$, $\mathrm{diag\{\boldsymbol{a}\}}$ is a diagonal matrix $\boldsymbol{A}$ of order $K$ for which $[\boldsymbol{A}]_{ii}=a_i~\forall i,~1\le i\le K$.}
%
%
\myitem{\ext{Ineuality overload:~}We overload inequality symbols to apply to real vectors of the same dimension in a componentwise fashion.
That is, for $\boldsymbol{a}, \boldsymbol{b}\in \mathbb{R}^{K\times 1}$, $\boldsymbol{a}<\boldsymbol{b}$ ($\boldsymbol{a} \le \boldsymbol{b}$) means $a_i < b_i$ ($a_i \le b_i$) $~\forall i,~1\le i\le K$.}
%
%
\myitem{\ext{Expectation:~}Finally, $\mathbb{E}\{\cdot\}$ stands for the statistical expectation operator and $\|\boldsymbol{a}\|_p$ is used to represent the p-norm of vector $\boldsymbol{a}$.}

\ifextended
\subsubsection{\color{olive}Organization}
\fi
In section II, we present the system model which consists of the frame structures and data rates, transfer and consumption of power to and in the WDs, as well as the representation of the UL and DL channels in the system.
\arx{In addition, a table of notations will be provided for future reference.}{}
We next solve what we call the forward problem;
that is, we derive an explicit expression for the UL WIT data rates of all WDs in terms of all parameters of the problem.
\arx{In order to do so, we need to study the UL and DL channels separately, and then equate the harvested power in the DL to the consumed power in the UL, tying together the two channels.}{}
Based on the results obtained, in section IV of the paper, these parameters are optimized such that the minimum UL WIT data rate among all WDs is maximized.
We then analyze the performance of the system in asymptotic regime when the number of antennas goes to infinity.
The next section provides simulation results to examine the truthfulness of our analytical findings as well as to offer the reader some interesting intuition.
The paper is finally concluded in section VII.


\section{System Model}
Consider a single-cell FDD WPCN consisting of a HAP equipped with $M$ antennas and $K$ single-antenna WDs
denoted by $WD_k,~k \in \boldsymbol{\mathcal{K}} = \{1,...,K\}$.
We assume that the WDs are sorted in an increasing order of their distances to the HAP.
We begin describing the model of this system with its frame structure which explains how different phases of information or power transmission are scheduled.
Then, we describe power transmission\arx{: the HAP sends a certain amount of power to each WD and the WDs harvest a portion of that power along with some additional interference power and utilize it to send data and feedback to the HAP.}{.}
Finally, the UL and DL communication channels are characterized in the last subsection.
\subsection{Frame Structure and Data Rates}

\begin{figure}[t]
\centering
\ifdouble
\includegraphics[scale=0.30]{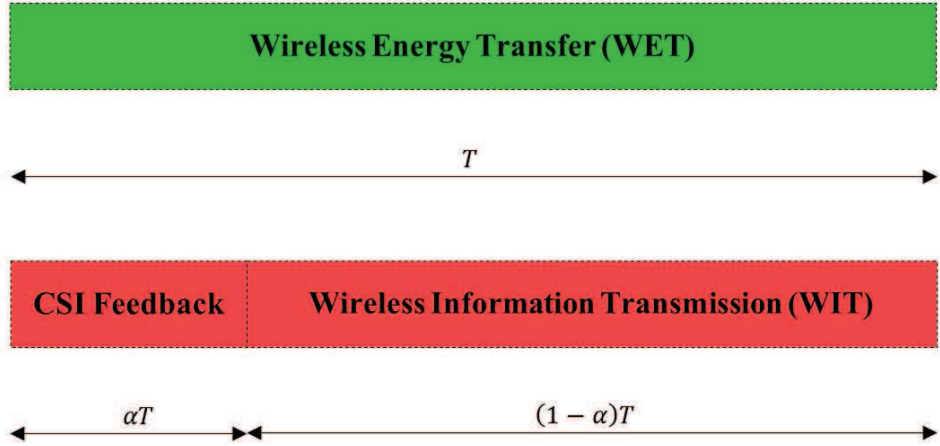}
\else
\includegraphics[scale=0.45]{frame.eps}
\fi

\caption{Frame structure. \emph{Top:} Downlink \emph{Bottom:} Uplink}
\label{fig.frame}
\end{figure}

The HAP spends the whole frame length, designated as
$T$, transmitting wireless energy to the WDs.
Using the harvested energy and through SDMA, all WDs, on the other hand, send data to the HAP in time $\left(1-\alpha \right)T$, and send CSI feedback in time $\alpha T$, where $\alpha$ is the \emph{CSI feedback phase time length ratio} and should satisfy $0\le \alpha \le 1$.
We represent the $k$-$th$ WD average UL data rate associated with the CSI feedback phase and the WIT phase by $r_{f,k}=\alpha r_{k}$ and $r_{w,k}=(1-\alpha)r_{k}$ respectively, where $r_k$ is the \emph{total UL data rate for the $k$-$th$ WD}. 
In the following sections, we will be switching back and forth between scalar notations, $r_{k,f}$, $r_{k,w}$, and $r_k$; and $\boldsymbol{r}^f$, $\boldsymbol{r}_w$, and $\boldsymbol{r}$ frequently.
It is further assumed that the HAP and all WDs are perfectly synchronized.
The UL and DL frame structures are shown in Fig. \myref[fig.frame].

\subsection{Power and Energy}
We assume the HAP has access to a permanent wired power supply but the WDs only receive wireless energy from the HAP through WET.
In particular, the HAP continuously transmits modulated power signals \cite{07081084} to the WDs with a \emph{maximum power spectral density} of $s_{\mathrm{max}}$ in a BW of $\beta B$, where $\beta$ is the \emph{DL BW ratio} and $B$ is the \emph{total BW}.
We assume the power is transmitted with maximum power spectral density $s_{\mathrm{max}}$ and so the total DL transmit power is $\beta s_{\mathrm{max}}B$.
\textblue{
However, the total DL transmit power $\beta s_{\mathrm{max}}B$ should not exceed \emph{maximum HAP power budget} $P_b>0$.
}
Moreover, the distribution of the harvested power among the WDs is not uniform.
In fact, as will be shown later, the HAP should send more power to WDs located farther from the HAP than those closer to it.
This is achieved through \emph{energy allocation weight vector} $\boldsymbol{\xiup}\in\mathbb{R}^{K\times1}$ which, because of the positivity of energy and the total DL power constraint, should satisfy $\boldsymbol{\xiup} \ge 0$, and $\|\boldsymbol{\xiup}\|_1=1$ respectively.

On the other hand, the expected harvested energy by wireless devices 1 through $K$ is represented by \emph{expected harvested energy vector} $\boldsymbol{\varepsilon}\in{\mathbb{R}}^{K\times1}$.
In our model, data transmission by WDs consumes all of the harvested power which is safe to assume for most low-power circuits.
In addition, because of the zero-forcing (ZF) receivers we will later employ at the HAP, the UL data transmission of the WDs cause no interference to each other and hence there is no reason not to transmit with maximum power \cite{01715541}\comm{06954434}.
As a result, the WDs transmit with constant \emph{UL power vector} $\boldsymbol{p}_{u}=T^{-1}\boldsymbol{\varepsilon}$ which means that the UL WIT and CSI feedback phases use equal powers.
As a result, the energy consumed in these phases becomes proportional to their time lengths, i.e. $\left(1-\alpha\right)$, and $\alpha$ respectively.

Note that we assume there is no power loss in the system, i.e. the AP and the WDs transmit and receive with absolute efficiency respectively.
In addition, our analysis only holds at the steady state where in each frame the amount of energy the WDs harvest is equal to the amount of energy they transmit.

\subsection{Uplink and Downlink Channels}
We assume no dominant line-of-sight propagation path between the HAP and the WDs exist and therefore we adapt the Rayleigh fading model for both the DL and UL channels.
In what follows we describe how the UL and DL channels are described using this model.

Assuming ${\boldsymbol{g}}_{u,i}~,~ i\in\boldsymbol{\mathcal{K}}$ are $M$-component \emph{UL channel vectors}, we define the \emph{UL channel matrix} by compiling
${\boldsymbol{g}}_{u,i}$
into a matrix
$\boldsymbol{G}_u={\left[\begin{array}{cccc}
{\boldsymbol{g}}_{u,1} & {\boldsymbol{g}}_{u,2} & \dots  & {\boldsymbol{g}}_{u,K} \end{array}
\right]}\in \mathbb{C}^{M \times K}$
where ${\boldsymbol{G}}_u$ is modeled as
\begin{equation*}
{\boldsymbol{G}}_u={\boldsymbol{H}}_u\mathrm{diag\left\{\boldsymbol{b}^{1/2}\right\}},
\end{equation*}
in which ${\boldsymbol{H}}_u\in {\mathbb{C}}^{M\times K}$ is the \emph{UL Rayleigh fading coefficient matrix} satisfying ${\left[{\boldsymbol{H}}_u\right]}_{mk}\sim \mathcal{C}\mathcal{N}\left(0,1\right)$, where $\mathcal{C}\mathcal{N}\left(\mu,\sigma^2\right)$ stands for circularly symmetric complex gaussian (CSCG) random variable with mean $\mu$ and variance $\sigma^2$, $\sim$ stands for ``distributed as'', and $\boldsymbol{b}$ is the $K$-component \emph{large-scale fading coefficient vector} each element of which, $b_k$, is assumed to be a known constant both at the HAP and $WD_k$  modeling channel path loss between the HAP and $WD_k$ \cite{07009979, 06375940}.
We assume the following long-term fading model holds
\def \eqLongTermFading{
\begin{equation}
\boldsymbol{b} = c_0d_0^\delta\boldsymbol{d}^{-\delta},
\label{eqLongTermFading}
\end{equation}
}
\eqLongTermFading
where $c_0$ is a constant representing attenuation at the reference distance $d_0$, $\delta>0$ is the \emph{pathloss exponent} and $\boldsymbol{d}$ is the $K$-component \emph{distance vector} of the WDs to the HAP.

Similarly, let us assume ${\boldsymbol{g}}_{d,i},\ i\in \boldsymbol{\mathcal{K}}$ are $M$-component \emph{DL channel vectors} and define the \emph{DL channel matrix} by compiling ${\boldsymbol{g}}_{d,i}$ into a matrix ${\boldsymbol{G}}_d\in\mathbb{C}^{M\times K}$ and model ${\boldsymbol{G}}_d$ as ${\boldsymbol{G}}_d = \boldsymbol{H}_d\mathrm{diag\{\boldsymbol{b}^{1/2}}\}$ where ${\boldsymbol{H}}_d\in {\mathbb{C}}^{M\times K}$ is the \emph{DL Rayleigh fading coefficient matrix} satisfying ${\left[{\boldsymbol{H}}_d\right]}_{mk}\sim \mathcal{C}\mathcal{N}\left(0,1\right)$ \cite{07009979}.

The DL channel vector ${\boldsymbol{g}}_{d,k}$ for $WD_k$ is estimated and then sent back to the HAP via CSI feedback.
We assume that the DL channel vector is estimated sufficiently well at each WD such that channel estimation error incurs negligible performance degradation.
The HAP receives the quantized, and normalized fed-back DL channel vector ${\tilde{\boldsymbol{g}}}_{d,k}$\textblue{ which we call \emph{fed-back DL channel vector} for brevity.}
For the sake of consistency of notation, these vectors are represented as columns of quantized, and normalized fed-back DL channel matrix ${\tilde{\boldsymbol{G}}}_{d}$ \textblue{which we call {fed-back DL channel matrix} for brevity.}

Lastly, we assume that the BW allocated to DL and UL channels are $\beta B$ and $(1-\beta)B$ respectively
and that both channels are quasi-static flat-fading, where $\boldsymbol{G}_d$, $\boldsymbol{G}_u$ and consequently $\tilde{\boldsymbol{G}}_d$ are constant during each block, but
can change from one block to another in accordance with the fading probability density function (PDF).
\arx{The latter assumption, called \textit{block fading}, is reasonable to make in systems with stationary nodes or systems where nodes move at walking speed.}{}

\ifarxiv
\begin{table}
\centering
\caption{\textblue{Table of Notations.}}
\label{table:notations}
\ifdouble
\resizebox{250pt}{!}{\begin{tabular}{ l l c }
\else
\resizebox{300pt}{!}{\begin{tabular}{ l l c }
\fi
\hline
Parameter & Description & Dimension\\ 
\hline
\hline
M        & Number of HAP antennas & $\mathbb{R}$   \\ \hline
K        & Number of WDs          & $\mathbb{R}$   \\ \hline
$\alpha$ & CSI feedback phase time length ratio & $\mathbb{R}$   \\ \hline
$\beta$  & CSI feedback phase time length ratio & $\mathbb{R}$   \\ \hline
$\boldsymbol{r}$  & Total UL data rate          & $\mathbb{R}^K$   \\ \hline
$\boldsymbol{r}_w$  & UL WIT data rate          & $\mathbb{R}^K$   \\ \hline
$\boldsymbol{r}_f$  & UL Feedback data rate          & $\mathbb{R}^K$   \\ \hline
$s_\mathrm{max}$  & maximum power spectral density & $\mathbb{R}$   \\ \hline
$B$      & Total Bandwidth        & $\mathbb{R}$   \\ \hline
$P_b$    & Maximum HAP power budget  & $\mathbb{R}$   \\ \hline
$\boldsymbol{\xiup}$    & energy allocation weight vector  & $\mathbb{R}^M$   \\ \hline
$\boldsymbol{p}_{u}$    &  UL power vector  & $\mathbb{R}^M$   \\ \hline
$\boldsymbol{\varepsilon}$    & expected harvested energy vector  & $\mathbb{R}^M$   \\ \hline
${\boldsymbol{g}}_{u,i}$    & UL channel vector  & $\mathbb{C}^M$   \\ \hline
${\boldsymbol{g}}_{d,i}$    & DL channel vector  & $\mathbb{C}^M$   \\ \hline
$\boldsymbol{G}_{u}$    & UL channel matrix  & $\mathbb{C}^{M\times K}$   \\ \hline
$\boldsymbol{G}_{d}$    & DL channel matrix  & $\mathbb{C}^{M\times K}$   \\ \hline
$\boldsymbol{H}_{u}$    & UL Rayleigh fading coefficient matrix  & $\mathbb{C}^{M\times K}$   \\ \hline
$\boldsymbol{H}_{d}$    & DL Rayleigh fading coefficient matrix  & $\mathbb{C}^{M\times K}$   \\ \hline
$\boldsymbol{b}$    & large-scale fading coefficient vector  & $\mathbb{C}^{K}$   \\ \hline
$\tilde{\boldsymbol{g}}_{d,i}$  & fed-back DL channel vector  & $\mathbb{C}^M$   \\ \hline
$\tilde{\boldsymbol{G}}_{d}$    & fed-back DL channel matrix  & $\mathbb{C}^{M\times K}$   \\ \hline
$c_0$    & constant attenuation at the reference distance $d_0$  & $\mathbb{R}$   \\ \hline
$\delta$    & pathloss exponent  & $\mathbb{R}$   \\ \hline

\end{tabular}}

\end{table}
\fi

\section{The Forward Problem}
In this section, we analyze the UL and DL channels to solve the forward problem of calculating the WIT data rate for every WD in term of the optimization varaiables.
In other words, we will derive an explicit formula for $\boldsymbol{r}_w(\alpha, \beta, \boldsymbol{\xiup})$.

\arx{Note that vector $\boldsymbol{r}_w$ is a UL parameter.
On the other hand, while $\alpha$ and $\beta$ are related to both DL and UL, $\boldsymbol{\xiup}$ is related to DL only.
As a result, to arrive at the desired formula, we will first examine the UL and DL channels separately, and then we will combine the results to arrive at the final explicit formula.}{}
\subsection{Uplink Transmission}
As mentioned earlier, the UL transmission consists of two phases, namely a \emph{CSI feedback phase} of length $\alpha T$ and a \emph{WIT phase} of length $(1-\alpha) T$ which we will analyze in this subsection.
In particular, we derive a formula for the total UL data rate of the $k$-$th$ WD and express the DL channel vector error incurred through feedback in terms of CSI feedback phase time length ratio $\alpha$.

\subsubsection{Wireless Information Transmission}

Let $\boldsymbol{y}$ be the $M$-component received complex baseband signal vector at the HAP in the WIT or CSI feedback phase 
\def \eqULSignal{
\begin{equation}
\boldsymbol{y}={\boldsymbol{G}}_u\left(\boldsymbol{p}_u^{1/2}\odot \boldsymbol{s}\right)+{\boldsymbol{n}}_u,
\label{eqULSignal}
\end{equation}
}
\eqULSignal
where $\boldsymbol{s}\sim \mathcal{C}\mathcal{N}\left(\boldsymbol{0}_K,\boldsymbol{I}_{K}\right)$ is the $K$-component \emph{UL information carrying signal vector}, and ${\boldsymbol{n}}_u\sim \mathcal{C}\mathcal{N}\left(\boldsymbol{0}_M,{\sigma }^2_{u,n}{\boldsymbol{I}}_{M}\right)$ is the $M$-component \emph{UL noise vector}, where $\sigma_{u,n}^2$ is the \emph{UL noise variance}.
A linear detector is used at the HAP to detect the signal transmitted by WDs.
Here, we use the ZF detector
\def \eqZFDetector{
\begin{equation}
\boldsymbol{A}={\boldsymbol{G}}_u{\left({\boldsymbol{G}}^H_u{\boldsymbol{G}}_u\right)}^{-1},
\label{eqZFDetector}
\end{equation}
}
\eqZFDetector
which yields $K$-component \emph{detected signal vector} $\boldsymbol{r}$ given by
\def \eqDetectedSignalVector{
\begin{equation}
\boldsymbol{r}={\boldsymbol{A}}^H{\boldsymbol{G}}_u\left(\boldsymbol{p}_u^{1/2}\odot \boldsymbol{s}\right)+{\boldsymbol{A}}^H{\boldsymbol{n}}_u.
\label{eqDetectedSignalVector}
\end{equation}
}
\eqDetectedSignalVector
Letting ${\boldsymbol{a}}_i$ denote the $i$-$th$ column of $\boldsymbol{A}$, the $k$-$th$ WD's detected signal, shown by $r_k$ can be expressed as
\def \eqDetectedSignalScalar{
\begin{equation}
r_k=\sqrt{p_{u,k}}{\boldsymbol{a}}^H_k{\boldsymbol{g}}_{u,k}s_k +\sum^K_{i=1,i\neq k}{\sqrt{p_{u,i}}{\boldsymbol{a}}^H_k{\boldsymbol{g}}_{u,i}s_i}+{\boldsymbol{a}}^H_k{\boldsymbol{n}}_u,
\label{eqDetectedSignalScalar}
\end{equation}
}
\eqDetectedSignalScalar
using which we can calculate $\gamma_k$, the signal-to-interference-plus-noise-ratio (SINR) for $WD_k$
\def \eqSINROne{
\begin{equation}
\gamma_k=\frac{p_{u,k}{\left|{\boldsymbol{a}}^H_k{\tilde{\boldsymbol{g}}}_{u,k}\right|}^{2}}     {{\sum^K_{i=1,i\mathrm{\neq }k}{p_{u,i}\left|{\boldsymbol{a}}^H_k{\tilde{\boldsymbol{g}}}_{u,i}\right|}}^{2}+\left|{\boldsymbol{a}}^H_k{\boldsymbol{a}}_k\right|\sigma^{2}_{u,n}},
\label{eqSINROne}
\end{equation}
}
\eqSINROne
and the achievable total UL data rate for $WD_k$ can therefore be written as $r_k = B\mathbb{E}\left\{{\log_2 \left(1+\gamma_k\right)}\right\}$, where $B$ is the UL BW.
A lower bound for the total UL data rate for $WD_k$ is $\tilde{r}_k$
\def \eqShannon{
\begin{equation}
r_k\ge {\tilde{r}}_k\equiv (1-\beta)B\log_2 \left(1+{\tilde{\gamma }}_k\right),
\label{eqShannon}
\end{equation}
}
\eqShannon
where
\def \eqSINRTwo{
\begin{equation}
{\tilde\gamma }_k^{-1}={\mathbb{E}\left\{\frac{{\sum^K_{i=1,i\neq k}{p_{u,i}\left|{\boldsymbol{a}}^H_k{\boldsymbol{g}}_{u,w,i}\right|}}^2+\left|{\boldsymbol{a}}^H_k{\boldsymbol{a}}_k\right|\sigma^{2}_{u,n}}   {p_{u,k}{\left|{\boldsymbol{a}}^H_k{\boldsymbol{g}}_{u,w,k}\right|}^2}\right\}}.
\label{eqSINRTwo}
\end{equation}
}
\eqSINRTwo
Following a procedure similar to that in \cite[lemma 4]{07009979}, we can simplify this expression to
\def \eqSINRThree{
\begin{equation}
{\tilde{\gamma }}_k=\frac{p_{u,k}\left(M-K\right)b_{k}}  {{\sigma }^2_{u,n}}, \quad M>K.
\label{eqSINRThree}
\end{equation}
}
\eqSINRThree

\arx{It should be emphasized that the analysis in this subsection applies to both phases in the UL transmission.}{}
In the rest of the paper, we will assume $M>K$ and use $\gamma$ and $r$ in lieu of diacritical characters $\tilde{\gamma}$ and $\tilde{r}$.
Therefore, using the definition for $\boldsymbol{r}_{w}$ and (\myref[eqShannon])
\def \eqDataRateWIT{
\begin{equation}
\boldsymbol{r}_{w}=(1-\alpha)(1-\beta)B\log_2(1+\boldsymbol{\gamma}).
\label{eqDataRateWIT}
\end{equation}
}
\eqDataRateWIT

\subsubsection{CSI Feedback}
In the CSI feedback phase, the directions of the estimated channel vectors are fed back to the HAP via CSI feedback.
This is done using a so-called \emph{codebook}, known both to the HAP and the WDs.
Consequently, the WDs only need to send the index of the closest code (vector) to the HAP, hence feeding-back the channel information.

Note that, generally, the optimal vector quantizer for this problem is not known.
One approach for creating the codebook is to choose all of the quantization vectors independently from the isotropic distribution on a unit sphere of M-dimensions \cite{Jindal10, 01715541}, \comm{also Jindal11}a technique referred to as \emph{Random Vector Quantization} (RVQ).
\arx{RVQ is easy to analyze and its performance is very close to optimal quantization \cite{01715541}.}{}

We assume that the number of feedback bits for $WD_k$ is given by $n_k=Tr_{f,k}$ and define the \emph{DL channel vector feedback quantization error} for $WD_k$ as
\def \eqFeedbackErrorOne{
\begin{equation}
\sigma_{u,f,k}^2 \equiv E_{\tilde{\boldsymbol{G}},\hat{\boldsymbol{G}}}\left\{\sin^2\left(\angle( \tilde{\boldsymbol{g}}_{d,k}, {\boldsymbol{g}_{d,k}})\right)\right\}.
\label{eqFeedbackErrorOne}
\end{equation}
}
\eqFeedbackErrorOne
In \cite{01715541} \comm{also LoveRef} it was shown that
\def \eqFeedbackErrorTwo{
\begin{equation}
\sigma_{u,f,k}^2 = 2^{n_k}\beta\left(2^{n_k},\frac{M}{M-1}\right) < 2^{-\frac{n_k}{M-1}},
\label{eqFeedbackErrorTwo}
\end{equation}
}
\eqFeedbackErrorTwo
where $\beta(\cdot)$ is the beta function defined by $\beta(p,q)=\frac{\Gamma(p)\Gamma(q)}{\Gamma(p+q)},$ where $\Gamma(\cdot)$ is the gamma function.
This upper error bound will be used in the DL section to derive the harvested energy formula.

\subsection{Downlink Transmission}

The DL transmission consists of a DL WET phase only where the DL transmission power $\beta s_{\mathrm{max}}$ is transferred via $M$-component \emph{beamforming vector} $\boldsymbol{w}$, where we assume $\|\boldsymbol{w}\|_2=1$.
This vector is used to adjust the energy transmit direction adaptively according to the instantaneous CSI of each frame \cite{06623072}.
WDs 1 through $K$ receive the complex baseband signals $z_k,~ k\in\boldsymbol{\mathcal{K}}$ where vector $\boldsymbol{z}\in \mathbb{R}^{K\times1}$ is expressed as

\def \eqDLSignal{
\begin{equation}
\boldsymbol{z}=\sqrt{B\beta s_{\mathrm{max}}}{\boldsymbol{G}}^H_{d}\boldsymbol{w}+\boldsymbol{n}_{d},
\label{eqDLSignal}
\end{equation}
}
\eqDLSignal
in which $\boldsymbol{n}_d \sim \mathcal{C}\mathcal{N}\left(\boldsymbol{0}_K\boldsymbol{0}_K^T, \sigma^2_{d,n}\boldsymbol{I}_k\right)$ is used to represent the WD noise vector whose elements we assume are negligible.
As a result, the expected harvested energy vector is
\def \eqEnergyVectorOne{
\begin{equation}
\boldsymbol{\varepsilon} = T B\beta s_{\mathrm{max}} \mathbb{E}_{\boldsymbol{G}_d,\hat{\boldsymbol{G}}_d }\left\{{\left|{\boldsymbol{G}}^H_{d}\boldsymbol{w}\right|}^2\right\}.
\label{eqEnergyVectorOne}
\end{equation}
}
\eqEnergyVectorOne
In words, the expected harvested energy by $WD_k$ is proportional to the square of the vector projection of the DL channel vector $\boldsymbol{g}_{d,k}$ onto the beamforming vector $\boldsymbol{w}$ \cite{07511170}.

We now need to find the \emph{pareto optimal energy beamformer} for this problem which maximizes the harvested energy for a given energy allocation weight vector.
This problem is formally defined as the following vector optimization problem
\def \prParetoOptimalBeamformer{
\begin{subequations}
\begin{align}
& & \underset{\boldsymbol{w}}{\text{maximize}}
& & & {\boldsymbol{\varepsilon}(\boldsymbol{w})},
& & & & \\
& & \text{subject to}
& & & \|\boldsymbol{w}\|_2=1.
\label{prParetoOptimalBeamformer}
\end{align}
\end{subequations}
}
\prParetoOptimalBeamformer
This is an optimization problem with a vector-valued objective function for which the set of achievable objective values does not have a maximum element, but rather a set of maximal elements, hence the name pareto optimal \cite{OptimizationBookRef}.
In lemma \myref[lem:pareto.optimal.beamformer] we obtain such a pareto optimal beamformer and derive a simple sufficient condition of its pareto optimality.

\def \eqParetoOptimalBeamformer{
\begin{equation}
\boldsymbol{w}={\tilde{\boldsymbol{G}}}_d{\boldsymbol{\xiup}^{1/2}},
\label{eqParetoOptimalBeamformer}
\end{equation}
}

\begin{lem}
\label{lem:pareto.optimal.beamformer}
The pareto optimal beamformer can be written as a linear combination of the normalized fed-back channel estimates
%
\eqParetoOptimalBeamformer
provided that each WD sends at least one feedback bit.
\end{lem}

\emph{Proof}: See Appendix \myref[app.pareto.optimal.beamformer]. 
\hfill $\blacksquare$

Equation (\myref[eqParetoOptimalBeamformer]) means that the allocated energy should be sent along the quantized, and normalized fed-back DL channel vectors of WDs.
Therefore, in the absence of any feedback error, the pareto optimal beamformer becomes a linear combination of the set of DL channel vectors.
Note that when the HAP intends to send energy to a single WD, for example wireless device 1, we have $\xiup_1=1, \xiup_{i}=0,~\forall i=2,...,K$, and the beamforming vector reduces to $\boldsymbol{w}=\boldsymbol{\tilde{g}}_{d,1}$, which maximizes the harvested energy for a single WD and is known as maximum ratio transmission (MRT) \cite{06623072}.
Therefore (\myref[eqParetoOptimalBeamformer]) is a direct extension of MRT.
\arx{We expect that the more power the HAP allocates to a DL channel vector, the more power the WD corresponding to that specific DL channel vector receives.
This is in fact true and is verified in lemma \myref[lem:EnergyVector] where we derive the amount of expected energy harvested by WDs.}{In lemma \myref[lem:EnergyVector] we derive the amount of expected energy harvested by WDs.}

\def \eqEnergyVectorTwo{
\begin{equation}
\boldsymbol{\varepsilon}=T B\beta s_{\mathrm{max}} \boldsymbol{b}\odot(\boldsymbol{M}\boldsymbol{\xiup}),
\label{eqEnergyVectorTwo}
\end{equation}
}
\begin{lem}
\label{lem:EnergyVector}
The expected harvested energy vector is given by
\eqEnergyVectorTwo
where $\boldsymbol{M}$ is the $K\times K$ \textit{mixing power matrix} for which $[\boldsymbol{M}]_{\{k\}\{k\}} = M \left(1-\sigma_{u,f,k}^2\right),~\forall k \in \boldsymbol{\mathcal{K}}$ and whose off-diagonal elements are all one.
\end{lem}

\emph{Proof}:
Following a similar procedure as that in \cite[lemma 2]{07009979} and using \cite{KayRef}
\ifdouble
\def \eqChannelOuterProd{
\begin{multline}
\mathbb{E}_{{\boldsymbol{G}}_d|{\tilde{\boldsymbol{G}}}_d}\left[{\boldsymbol{g}}_{d,i}{\boldsymbol{g}}^H_{d,j}\right]\\
=b_i\left\{ \begin{array}{lc}
\boldsymbol{0}_M\boldsymbol{0}_M^T, & i\neq j \\ 
{\sigma }^2_{u,f,i}{\boldsymbol{I}}_M+\left(1-{\sigma }^2_{u,f,i}\right)M{\tilde{\boldsymbol{g}}}_{d,i}{\tilde{\boldsymbol{g}}}^H_{d,i}, & i=j \end{array}
\right.
\label{eqChannelOuterProd}
\end{multline}
}
\else
\def \eqChannelOuterProd{
\begin{equation}
\mathbb{E}_{{\boldsymbol{G}}_d|{\tilde{\boldsymbol{G}}}_d}\left[{\boldsymbol{g}}_{d,i}{\boldsymbol{g}}^H_{d,j}\right]
=b_i\left\{ \begin{array}{lc}
\boldsymbol{0}_M\boldsymbol{0}_M^T, & i\neq j \\ 
{\sigma }^2_{u,f,i}{\boldsymbol{I}}_M+\left(1-{\sigma }^2_{u,f,i}\right)M{\tilde{\boldsymbol{g}}}_{d,i}{\tilde{\boldsymbol{g}}}^H_{d,i}, & i=j \end{array}
\right.
\label{eqChannelOuterProd}
\end{equation}
}
\fi
\eqChannelOuterProd
we can derive
\def \eqEnergyScalar{
\begin{equation}
{\varepsilon }_k = T B\beta s_{\mathrm{max}} b_k\left\{M\left(1-{\sigma }^2_{u,f,k}\right)\xiup_k+\sum_{i=1,i\neq k}^{K}{\xiup}_i  \right\},
\label{eqEnergyScalar}
\end{equation}
}
\eqEnergyScalar
which can be compactly written as (\myref[eqEnergyVectorTwo]).
\hfill$\blacksquare$

Two observations are in order.
First, from the structure of the mixing power matrix $\boldsymbol{M}$, not to be confused with the number of HAP antennas $M$, we realize that, assuming DL channel vector feedback quantization error for a particular WD is negligible, the energy harvested from the beam aimed at this WD is multiplied by the number of HAP antennas $M$, while the power harvested from beams aimed at other WDs is multiplied by one.
We call the first and the second term the \emph{beamed energy} and the \emph{interference energy} respectively.
It is because of this multiplication factor $M$ that the HAP can control the distribution of power among the WDs.
Second, whether the multiplication factor $M$ is effective depends upon the amount of DL channel vector feedback quantization error as a large error makes the diagonal elements diminish.
In the extreme case, when $n_k=0$, the beamforming vector $ \boldsymbol{w}$ becomes essentially random with respect to the channel vector $\boldsymbol{g}_{d,k}$ and hence the diagonal element for $WD_k$ becomes 1.
This is verified mathematically if we note that from (\myref[eqFeedbackErrorTwo]), the expected value of the DL channel vector feedback quantization error squared $\sigma_{u,f,k}^2$ is equal to $1-1/M$ at $n_k=0$.
When the feedback is eliminated for all the WDs, the harvested energy becomes $\boldsymbol{\varepsilon} = TB \beta s_{\mathrm{max}} \boldsymbol{b}$, at which point the HAP totally fails to control the distribution of energy among the WDs.

\subsection{Solution to the Forward Problem}
So far, we have analyzed the DL and UL channels separately.
Yet, these channels are coupled through energy.
In this subsection we combine the DL and UL formulas to obtain an explicit expression of the total UL data rate for every WD in terms of our decision variables $\alpha$, $\beta$ and $\boldsymbol{\xiup}$.
%
%

Using (\myref[eqSINRThree]) and (\myref[eqEnergyVectorTwo]), the SINR vector can be written as
\def \eqLambdaOne{
\begin{equation}
\boldsymbol{\lambda} = B\beta s_{\mathrm{max}}\frac{\left(M-K\right)}{\sigma_{u,n}^2} \boldsymbol{b}^2 \odot (\boldsymbol{M}\boldsymbol{\xiup}).
\label{eqLambdaOne}
\end{equation}
}
\eqLambdaOne
Let us decompose $\boldsymbol{M}$ to arrive at a more intuitive formula
\def \eqMDecomposition{
\begin{equation}
\boldsymbol{M} = \boldsymbol{M}^\prime -\boldsymbol{M}^{\prime\prime},
\label{eqMDecomposition}
\end{equation}
}
\eqMDecomposition
where $\boldsymbol{M}^\prime$s diagonal and off-diagonal elements are $M$ and one respectively and $\boldsymbol{M}^{\prime\prime}=M\mathrm{diag}\{\boldsymbol{\sigma}_{u,f}^2\}$.
Then $\boldsymbol{\gamma}$ can be written as $\boldsymbol{\gamma} = \boldsymbol{\gamma}^{\mathrm{max}}-\boldsymbol{\gamma}^{\mathrm{loss}}$ where
\def \eqLambdaMax{
\begin{equation}
\boldsymbol{\gamma}^\mathrm{max} = B\beta s_{\mathrm{max}}\frac{\left(M-K\right)}{{\sigma }^2_{u,n}} \boldsymbol{b}^2 \odot \left(\boldsymbol{M}^\prime\boldsymbol{\xiup}\right),
\label{eqLambdaMax}
\end{equation}
}
\eqLambdaMax
\def \eqLambdaLossOne{
\begin{equation}
\boldsymbol{\gamma}^\mathrm{loss} = B\beta s_{\mathrm{max}}\frac{\left(M-K\right)}{\sigma^2_{u,n}} \boldsymbol{b}^2 \odot \left(\boldsymbol{M}^{\prime\prime}\boldsymbol{\xiup}\right).
\label{eqLambdaLossOne}
\end{equation}
}
\eqLambdaLossOne
We can express $\boldsymbol{\gamma}^{\mathrm{loss}}$ as
\def \eqLambdaLossTwo{
\begin{equation}
\boldsymbol{\gamma}^{\mathrm{loss}} = \boldsymbol{\gamma}^{\mathrm{maxloss}}\odot\boldsymbol{\sigma}_{u,f}^2,
\label{eqLambdaLossTwo}
\end{equation}
}
\eqLambdaLossTwo
where
\def \eqLambdaMaxloss{
\begin{align}
\boldsymbol{\gamma}^\mathrm{maxloss} = B\beta s_{\mathrm{max}}\frac{M\left(M-K\right)}{\sigma^2_{u,n}} \boldsymbol{b}^2 \odot \boldsymbol{\xiup}.
\label{eqLambdaMaxloss}
\end{align}
}
\eqLambdaMaxloss
Combining equations (\myref[eqShannon]), (\myref[eqFeedbackErrorTwo]), and (\myref[eqLambdaLossTwo]) we get
\def \eqFundamentalOne{
\begin{equation}
\boldsymbol{r} = (1-\beta)B \log_2 \left(1 + \boldsymbol{\gamma}^\mathrm{max} - \boldsymbol{\gamma}^\mathrm{maxloss} \odot 2^{-\alpha \left(\frac{T\boldsymbol{r}}{M-1}\right)} \right).
\label{eqFundamentalOne}
\end{equation}
}
\eqFundamentalOne
This equation means that the overall effect of CSI feedback is to reduce the SINR by $\boldsymbol{\gamma}^\mathrm{maxloss} \odot 2^{-\alpha \left(\frac{T\boldsymbol{r}}{M-1}\right)}$ which is a decreasing function of $\alpha$.
This equation, however, is implicit, because the WIT data rate loss for a WD is affected by the number of CSI feedback bits that is being transmitted which, itself depends upon the total UL data rate for that particular WD.
In lemma \myref[lem:UL.data.rate], we derive an explicit formula for the UL data rate loss of every WD, from which the WIT data rate may be easily calculated.

\def \eqFeedbackErrorThree{
\begin{equation}
{\sigma}_{u,f,k}^2 = \frac{1+\gamma^{\mathrm{max}}_k}{{\left(1+\gamma^{\mathrm{max}}_k\right)}^{1+\alpha \left(\frac{TB}{M-1}\right)}-\alpha \left(\frac{TB}{M-1}\right)\gamma^{\mathrm{maxloss}}_k}.
\label{eqFeedbackErrorThree}
\end{equation}
}
\begin{lem}
\label{lem:UL.data.rate}
The DL channel vector feedback quantization error for $WD_k$ is equal to
\eqFeedbackErrorThree
\end{lem}

\emph{Proof}: See Appendix \myref[app.UL.data.rate].
\hfill $\blacksquare$

Using (\myref[eqDataRateWIT]), (\myref[eqLambdaLossTwo]), and (\myref[eqFeedbackErrorThree]) we can get an explicit formula relating the WIT data rate of every WD in terms of our decision variables
\ifdouble
\def \eqDataRateWITThree{
\begin{multline}
r_{w,k} = (1-\alpha)(1-\beta)B\\
 \log_2 \left(1 + \gamma_k^\mathrm{max} -  \frac{\gamma_k^\mathrm{loss}\left(1+\gamma^{\mathrm{max}}_k\right)}{{\left(1+\gamma^{\mathrm{max}}_k\right)}^{1+\alpha \left(\frac{TB}{M-1}\right)}-\alpha \left(\frac{TB}{M-1}\right)\gamma^{\mathrm{maxloss}}_k} \right).
\label{eqDataRateWITThree}
\end{multline}
}
\else
\def \eqDataRateWITThree{
\begin{equation}
r_{w,k} = (1-\alpha)(1-\beta)B \log_2 \left(1 + \gamma_k^\mathrm{max} -  \frac{\gamma_k^\mathrm{loss}\left(1+\gamma^{\mathrm{max}}_k\right)}{{\left(1+\gamma^{\mathrm{max}}_k\right)}^{1+\alpha \left(\frac{TB}{M-1}\right)}-\alpha \left(\frac{TB}{M-1}\right)\gamma^{\mathrm{maxloss}}_k} \right).
\label{eqDataRateWITThree}
\end{equation}
}
\fi
\eqDataRateWITThree
We can now proceed to optimization.

\section{Throughput Optimization}
To optimize the throughput and improve the WIT data rates while attaining fairness, we propose to maximize the minimum WIT data rate among all WDs \comm{07386616}; that is, to solve the following optimization problem
\def \prOptimizationOne{
\begin{subequations}
\label{prOptimizationOne}
\begin{align}
& & \underset{\alpha, \beta, \boldsymbol{\xiup}}{\text{maximize}}
& & & {\mathop{\mathrm{min}}_{k\in\boldsymbol{\mathcal{K}}} r_{w,k}(\alpha, \beta, \boldsymbol{\xiup})}\label{eqXiPositivityCostFunc},
& & & & \\
& & \text{subject to}
& & & \boldsymbol{\xiup}\ge 0,\label{eqXiPositivityConstraintEq}\\
& &
& & & \|\boldsymbol{\xiup}\|_1=1,\label{eq:Xi_norm_constraint_eq}\\
& &
& & & 0\le\beta \le 1,\label{UeqBetaLimitEq}\\
& &
& & & \beta B s_{\mathrm{max}}<P_b,\label{eqBetaConstraintEq}\\
& &
& & & 0\le\alpha \le 1.\label{eqAlphaConstraintEq}
\end{align}
\end{subequations}
}
\prOptimizationOne
As can be seen, the optimization variables are CSI feedback phase time length ratio $\alpha$, DL BW ratio $\beta$, and energy allocation weight vector $\boldsymbol{\xiup}$ which are interrelated as follows:
The amount of energy harvested by every WD in the WET phase is basically controlled by energy allocation weight vector $\boldsymbol{\xiup}$ and DL BW ratio $\beta$ but is also affected by CSI accuracy.
On the other hand, CSI accuracy is dependent upon the DL channel vector feedback quantization error which is, in turn, determined by feedback data rate $r_{f,k}$.
The feedback and WIT data rates for every WD depend on the length of the feedback and WIT phases, i.e. $\alpha T$ and $(1-\alpha)T$ respectively, the UL channel BW $(1-\beta)B$, and the corresponding SINR at the HAP.
Finally, the SINR at the HAP is related to the harvested energy by the WD in question.

%
%
%

\textblue{Solving (\myref[prOptimizationOne]) efficiently depends upon the fact that whether or not the problem is convex, which, in turn, requires (\myref[eqXiPositivityCostFunc]) to be a concave function.}
Proving the concavity of (\myref[eqDataRateWITThree]), however, is difficult.
Instead, in order to prove the existence and uniqueness of the solution of (\myref[prOptimizationOne]), we use (\myref[eqFundamentalOne]) and assume it is solved recursively and therefore the data rate at the previous recursion is constant; making (\myref[eqFundamentalOne]) explicit at each iteration.
Then, we prove the convexity of the WIT data rate equation at each recursion and show the convergence of recursions through simulations.
The proof for concavity of (\myref[eqFundamentalOne]) is given in the following lemma.

\begin{lem}
\label{lem:convexity1}
When solved recursively, (\myref[eqFundamentalOne]) is concave with recpect to $\alpha$, $\beta$, and $\boldsymbol{\xiup}$.
\end{lem}

\emph{Proof}:
Using (\myref[eqDataRateWIT]) and (\myref[eqLambdaOne]) 
\def \eqDataRateWITTwo{
\begin{equation}
\boldsymbol{r}_w = B (1-\alpha)(1-\beta) \log_2 \left(1+ B\beta s_{\mathrm{max}}\frac{\left(M-K\right)}{{\sigma }^2_{u,n}} \boldsymbol{b}^2 \odot (\boldsymbol{M}\boldsymbol{\xiup})  \right).
\label{eqDataRateWITTwo}
\end{equation}
}
\eqDataRateWITTwo
Note that
\def \eqMDiag{
\begin{equation}
\mathrm{diag}\left\{\boldsymbol{M}\right\} = M \left(\boldsymbol{1}-2^{-\frac{1}{M-1}\boldsymbol{n}}\right)=M\left(\boldsymbol{1}-2^{-\frac{\alpha T}{M-1}\boldsymbol{r}}\right),
\label{eqMDiag}
\end{equation}
}
\eqMDiag
therefore, assuming $\boldsymbol{r}$ is constant, $\boldsymbol{M}$ is a non-negative concave function of $\alpha$.
This means that $B\beta s_{\mathrm{max}}\frac{\left(M-K\right)}{{\sigma }^2_{u,n}} \boldsymbol{b}^2 \odot (\boldsymbol{M}\boldsymbol{\xiup})$ is a product of non-negative affine with concave functions.
As a result, it is a log concave function.
In addition, it can be shown that $\log_2\left(1+B\beta s_{\mathrm{max}}\frac{\left(M-K\right)}{{\sigma }^2_{u,n}} \boldsymbol{b}^2 \odot (\boldsymbol{M}\boldsymbol{\xiup})\right)$ and therefore its product with affine functions; that is $(1-\alpha)(1-\beta)\log_2\left(1+B\beta s_{\mathrm{max}}\frac{\left(M-K\right)}{{\sigma }^2_{u,n}} \boldsymbol{b}^2 \odot (\boldsymbol{M}\boldsymbol{\xiup})\right)$ is log-concave w.r.t. $\alpha$, $\boldsymbol{\xiup}$, and $\beta$ too.
\hfill $\blacksquare$

In what follows we proceed to find the optimal values of our decision variables one by one.

\subsection{Energy Allocation Weight Vector}
As distances of the WDs from the HAP can be widely different, DL signal attenuation and hence the harvested power varies greatly among different WDs.
Moreover, due to the UL signal attenuation, farther WDs from the HAP have to transmit with greater power so that the received signal at the HAP has the same SINR, a problem referred to as the ``double near-far'' \cite{07009979}, or ``doubly near-far'' effect \cite{07462480, 06678102}.
Thus, the energy allocation weight vector needs to be chosen in such a way as to cancel out this problem.

Suppose that we can partition the WD index set $\boldsymbol{\mathcal{K}}$ into two sets $\boldsymbol{\mathcal{K}}^f$ and $\boldsymbol{\mathcal{K}}^u$ where $\boldsymbol{\mathcal{K}}^f \cap \boldsymbol{\mathcal{K}}^u = \varnothing$ and $\boldsymbol{\mathcal{K}}^f \cup \boldsymbol{\mathcal{K}}^u = \boldsymbol{\mathcal{K}}$.
In theorem \myref[thm:OptimalXiup], the optimal energy allocation weight vector for this problem is derived.
Before that, however, lemma \myref[lem:ZeroXiup] characterizes $WD_k,~k \in \boldsymbol{\mathcal{K}}^u$.

\def \eqSINREquality{
\begin{equation}
B\beta s_{\mathrm{max}}\frac{\left(M-K\right)}{\sigma^2_{u,n}} \boldsymbol{b}^2 \odot (\boldsymbol{M}\boldsymbol{\xiup}) = \gamma_c \boldsymbol{1},
\label{eqSINREquality}
\end{equation}
}
\begin{lem}
\label{lem:ZeroXiup}
\textblue{The only case where the WIT data-rate for a WD is different from the others is when its energy allocation weight coefficient is zero.}
\end{lem}

\emph{Proof}:
Forcing the WIT data-rates for all WDs to be equal demands that the SINR at the HAP be the same for every WD 
\eqSINREquality
where $\gamma_c$ is the common SINR value.
Note, however, that this equation neglects the non-negativity of $\boldsymbol{\xiup}$.
The fact that some elements of $\boldsymbol{\xiup}$ should be negative to ensure fairness means that fairness cannot be achieved for such WDs.
The best choice, then, is to set the energy allocation coefficient of such WDs to zero.
\hfill $\blacksquare$

%
\def \eqKSet{
\begin{subequations}
\label{eqKSet}
\begin{align}
\boldsymbol{\mathcal{K}}_m^f & \subset\boldsymbol{\mathcal{K}}:\quad [\xiup]_{\boldsymbol{\mathcal{K}}_m^f} > 0 ,\label{eq:K_f_set}\\
\boldsymbol{\mathcal{K}}_n^u & \subset\boldsymbol{\mathcal{K}}: \quad [\xiup]_{\boldsymbol{\mathcal{K}}_n^u} = 0 ,\label{eq:K_u_set}
\end{align}
\end{subequations}
}
\textblue{
Based on this lemma, we can define sets $\boldsymbol{\mathcal{K}}^f$ and $\boldsymbol{\mathcal{K}}^u$ as follows
\eqKSet
where superscripts $f$ and $u$ stand for fair and unfair respectively.}
Assuming sets $\boldsymbol{\mathcal{K}}^f$ and $\boldsymbol{\mathcal{K}}^u$ are known, we now proceed to calculate the optimal energy allocation weight vector

\def \eqOptimalXiupOne{
\begin{subequations}
\label{eqOptimalXiupOne}
\begin{align}
\boldsymbol{\xiup}_{\boldsymbol{\mathcal{K}}^f} &= \|\boldsymbol{M}^{-1}_{\boldsymbol{\mathcal{K}}^f,\boldsymbol{\mathcal{K}}^f}\boldsymbol{b}^{-2}_{\boldsymbol{\mathcal{K}}^f}\|_1^{-1} \boldsymbol{M}^{-1}_{\boldsymbol{\mathcal{K}}^f,\boldsymbol{\mathcal{K}}^f}\boldsymbol{b}^{-2}_{\boldsymbol{\mathcal{K}}^f}, \label{eqOptimalXiupfair}\\
\boldsymbol{\xiup}_{\boldsymbol{\mathcal{K}}^u} &= \boldsymbol{0}. \label{eqOptimalXiupunfair}
\end{align}
\end{subequations}
}
\begin{thm}
\label{thm:OptimalXiup}
The optimal energy allocation weight vector $\boldsymbol{\mathrm{\xiup}}$ is given by
\eqOptimalXiupOne

\end{thm}

\emph{Proof}:
In lemma \myref[lem:ZeroXiup], we explained why the energy allocation vector of $WD_k, k\in \boldsymbol{\mathcal{K}}^u$ is set to zero.
On the other hand, for every $WD_k, k\in \boldsymbol{\mathcal{K}}^f$, the WIT data rate and as a result, the SINR should be the same at the HAP
\def \eqSINREquality{
\begin{equation}
B\beta s_{\mathrm{max}}\frac{\left(M-K\right)}{{\sigma }^2_{u,n}} \boldsymbol{b}^2_{\boldsymbol{\mathcal{K}}^f} \odot (\boldsymbol{M}_{\boldsymbol{\mathcal{K}}^f,\boldsymbol{\mathcal{K}}^f}\boldsymbol{\xiup}_{\boldsymbol{\mathcal{K}}^f}) = \gamma_c \boldsymbol{1},
\label{eqSINREquality}
\end{equation}
}
\eqSINREquality
where $\gamma_c$ is the common SINR whose value is determined upon normalization of $\boldsymbol{\xiup}$ vector.
The solution to this equation is (\myref[eqOptimalXiupfair]).
{Note that the condition for pareto-optimality of the beamformer (that is, each WD feeding back at least one bit) ensures invertibility of $\boldsymbol{M}$ and its submatrices.
\hfill $\blacksquare$

The procedure by which $\boldsymbol{\mathcal{K}}^f$ and $\boldsymbol{\mathcal{K}}^u$ are determined is as follows.
We begin by setting $\boldsymbol{\mathcal{K}}^f = \boldsymbol{\mathcal{K}}$, and $\boldsymbol{\mathcal{K}}^u = \varnothing$.
After computing the energy allocation weight vector, some of its elements may be negative.
If this is the case, then their indices should be added to $\boldsymbol{\mathcal{K}}^u$ and excluded from $\boldsymbol{\mathcal{K}}^f$ and the vector should be recomputed.
This process is repeated until the resulting energy allocation weight vector is non-negative.

\def \eqOptimizationTwo{
\begin{subequations}
\label{eqOptimizationTwo}
\begin{align}
& & \underset{\alpha,\beta}{\text{maximize}}
& & & r_{w,K}(\alpha,\beta, \boldsymbol{\xiup}^*),
& & & & \\
& & \text{subject to}
& & & (\myref[eqXiPositivityConstraintEq])-(\myref[eqBetaConstraintEq]).
\end{align}
\end{subequations}
}
According to (\myref[eqLongTermFading]), as the distance of $WD_k, k\in\boldsymbol{\mathcal{K}}^f$ to the HAP decreases, the large-scale fading coefficient $b_k$ is increased.
In a network having more than one WD, this, according to (\myref[eqOptimalXiupfair]), leads to a decrease in the corresponding energy allocation weight coefficient $\xiup_k$.
\color{black}
As $WD_k$ moves even closer to the HAP, $\xiup_k$ reaches zero, from which point onward $k\in \mathcal{K}^u$ and moving the device closer to the HAP has no effect on $\xiup_k$.
On the other hand, doing so will further increase $b_k$, which, according to (\myref[eqDataRateWITTwo]) increases $r_{w,k}$.
\arx{Qualitatively, bringing $WD_k$ closer to the HAP without changing $\boldsymbol{\xiup}$ will result $WD_k$ to receive more power than needed which gives rise to a higher maximum WIT data rate than other WDs.}{}

According to these definitions, (\myref[eqLambdaMax]), (\myref[eqLambdaLossOne]), (\myref[eqFundamentalOne]), and assuming $\boldsymbol{\sigma}_{u,f}^2 < \frac{1}{M}\boldsymbol{1}$
\ext{(This condition makes sure the diagonal elements of matrix $\boldsymbol{M}$ are positive.)}{}
\def \eqRMaxKSet{
\begin{subequations}
\label{eqRMaxKSet}
\begin{align}
r_k &= r_c , \quad k \in \boldsymbol{\mathcal{K}}^f,\label{eq:R_max_K_f_set}\\
r_k & > r_c , \quad \forall k \in \boldsymbol{\mathcal{K}}^u,
\label{eq:R_max_K_u_set}
\end{align}
\end{subequations}
}
\eqRMaxKSet
where $r_c$ is the common data rate in the fair region.
This can be justified because, in addition to the beamed energy controlled by $\boldsymbol{\xiup}$, every WD receives some interference energy as well.
At very low distances, the interference energy a WD receives alone might be sufficient or even more than sufficient to power its UL transmission so as to achieve the desired UL WIT data rate.
Hence the corresponding energy allocation weight coefficient becomes zero at such distances.

Geometrically, as shown in Fig. \myref[figFairnessFig], the area around the HAP can be divided into two regions: the \emph{unfair region} defined by $0<d \le r_f$ and the \emph{fair region} defined by $d > r_f$ where $d$ is the distance to the HAP and $r_f$ is the \emph{fairness radius} defined by the following theorem.
\def \eqFairnessRadiusOne{
\begin{equation}
r_f = \sqrt[2\delta]{\frac{\boldsymbol{v}^T \boldsymbol{d}^{2\delta}}{1+\boldsymbol{1}^T \boldsymbol{v}}}.
\label{eqFairnessRadiusOne}
\end{equation}
} 
\begin{thm}
\label{lem:fairness_radius}
The fairness radius is given by
\eqFairnessRadiusOne
\end{thm}

\emph{Proof}: 
The fairness radius is defined to be the distance at which the interference energy is exactly equal to the energy needed for the WDs placed at this distance to achieve the intended UL WIT data rate.
In order to calculate this distance, we first simplify $\boldsymbol{M}^{-1}$ using the Sherman-Morrison formula
\def \eqShermanMorrison{
\begin{equation*}
\boldsymbol{M}^{-1}
= \mathrm{diag}\left\{\boldsymbol{v}\right\} - \frac{1}{1+\boldsymbol{1}^T \boldsymbol{v} }\boldsymbol{v}\boldsymbol{v}^T,
\end{equation*}
}
\eqShermanMorrison
where $\boldsymbol{v}
= \left[(M-1)\boldsymbol{1}-M\boldsymbol{\sigma}_{u,f}^2\right]^{-1}$.
Next, we set the $k$-$th$ element of $\boldsymbol{\xiup}$ in (\myref[eqOptimalXiupfair]) to zero
\begin{equation*}
\left\{v_k\boldsymbol{e}_k^T - \frac{v_k}{1+\boldsymbol{1}^T \boldsymbol{v} } \boldsymbol{v}^T \right\} \boldsymbol{b}^{-2} = 0.
\end{equation*}
Simplification gives
\begin{equation*}
b_k^{-1} = \sqrt{\frac{\boldsymbol{v}^T \boldsymbol{b}^{-2}}{1+\boldsymbol{1}^T \boldsymbol{v}}}.
\end{equation*}
Substituting the path loss model for $\boldsymbol{b}$, (\myref[eqFairnessRadiusOne]) results.
\hfill $\blacksquare$

Note that, for a given system, the fairness radius is not fixed and depends on the distances of the WDs to the HAP.
\arx{Assuming the availability of full CSI at the HAP, $\boldsymbol{v} = (M-1)^{-1}\boldsymbol{1}$ and therefore
\begin{equation*}
r_f \approx (M+K-1)^\frac{-1}{2\delta}{\left\| \boldsymbol{d} \right\|}_{2\delta}.
\end{equation*}
Since $\delta \ge 2$, this equation may further be approximated by
\def \eqFairnessRadiusTwo{
\begin{equation}
r_f \approx (M+K-1)^\frac{-1}{2\delta}{\left\| \boldsymbol{d} \right\|}_{\infty}.
\label{eqFairnessRadiusTwo}
\end{equation}
}
\eqFairnessRadiusTwo

This formula simply means that the fairness radius roughly only depends on the maximum distance of the WDs to the HAP.
The closer the farthest WD is to the HAP, the lower this radius becomes and vice versa.}{}

\begin{figure}[t]
\ifdouble
\includegraphics[trim=90 210 540 170,clip,width=\textwidth]{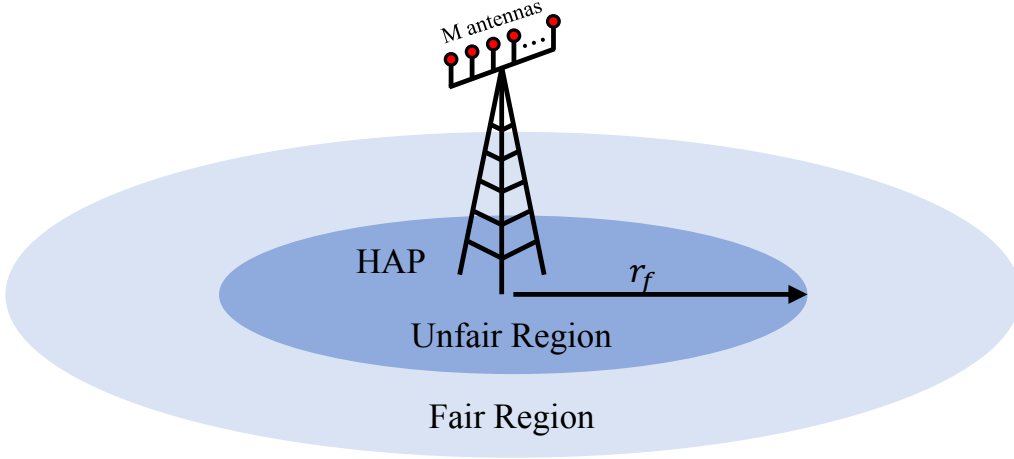}
\else
\includegraphics[trim=90 210 540 170,clip,scale = 1.2]{fairness.eps}
\fi
\centering
\caption{The area around the HAP is divided into a fair and an unfair region defined by the fairness radius $r_f$.}
\label{figFairnessFig}
\end{figure}

\subsection{Downlink Bandwidth Ratio}
\textblue{Now that the optimum $\boldsymbol{\xiup}$ for a particular choice of $\alpha$ and $\beta$ has been found, we can use this information to simplify the optimization problem (\myref[prOptimizationOne]})
\begin{lem}
\label{lem:optimization_problem4}
Having computed $\boldsymbol{\xiup}^*$, the optimal $\alpha$ and $\beta$ can be calculated from the following simplified optimization problem
\eqOptimizationTwo
\end{lem}

\emph{Proof:}
From (\myref[prOptimizationOne]), $\alpha$ and $\beta$ are the maximizers of $\mathop{\mathrm{min}}_{k\in\boldsymbol{\mathcal{K}}} r_{w,k}(\alpha,\beta,\boldsymbol{\xiup})$.
Yet, assuming $\boldsymbol{\xiup}=\boldsymbol{\xiup}^*$ and considering (\myref[eqRMaxKSet]),
\begin{equation*}
\mathop{\mathrm{min}}_{k\in\boldsymbol{\mathcal{K}}} r_{w,k}(\alpha,\beta,\boldsymbol{\xiup}) = \mathop{\mathrm{min}}_{k\in\boldsymbol{\mathcal{K}}^f} r_{w,k}(\alpha,\beta,\boldsymbol{\xiup}^*).
\end{equation*}
On the other hand, it is clear that the fair region always includes the farthest WD from the HAP, i.e. $K \in \boldsymbol{\mathcal{K}}^f$ which completes the proof.
\hfill $\blacksquare$

Increasing the DL BW by increasing DL BW ratio $\beta$ increases the harvested power by the WDs which leads to a higher SNR at the HAP and ultimately a higher WIT data rate.
Nevertheless, increasing the DL BW at the same time decreases the UL BW which directly decreases the WIT data rate.
Hence, an optimal value for the DL BW ratio exists which we will subsequently find.
%
%

In the following theorem, to show the dependence on $\beta$ explicitly, instead of $\gamma_k$, $\gamma_k^{\mathrm{max}}$, and $\gamma_k^{\mathrm{maxloss}}$, we will use $\bar{\gamma}_k = \gamma_k / \beta$, $\bar{\gamma}_k^\mathrm{max}=\gamma_k^{\mathrm{max}}/\beta$, and $\bar{\gamma}_k^{\mathrm{maxloss}}=\gamma_k^{\mathrm{maxloss}}/\beta$ respectively.

\def \eqOptimalBeta{
\begin{equation}
\beta(\boldsymbol{\xiup}) = \min\left\{\frac{\bar{\gamma }^{\mathrm{max}}_K+1}{\bar{\gamma }^{\mathrm{max}}_K W_0\left(e\left(\bar{\gamma }^{\mathrm{max}}_K+1\right)\right)}-\frac{1}{\bar{\gamma }^{\mathrm{max}}_K}, \frac{P_b}{Bs_{\mathrm{max}}} \right\},
\label{eqOptimalBeta}
\end{equation}
}
\begin{thm}
\label{thm:optimal_beta}
The optimal DL BW ratio is given by
\eqOptimalBeta
where $W_0\left(\cdot\right)$ is the principal branch of the Lambert W function.
\end{thm}

\emph{Proof}:
Taking the derivative of (\myref[eqShannon]) for $k=K$ with respect to $\beta$, we get
\begin{equation*}
\ln 2\ \frac{\partial r_K}{\partial \beta }=\frac{\bar{\gamma}_K B\left(1-\beta \right)}{1+\beta \bar{\gamma}_K}-B{\ln \left(1+\beta \bar{\gamma}_K\right)\ }.
\end{equation*}
Setting $\frac{\partial r_K}{\partial \beta } = 0$ yields \ifextended{\color{olive}$\frac{\bar{\gamma}_K\left(1-\beta \right)}{1+\beta \bar{\gamma}_K}={\ln \left(1+\beta \bar{\gamma}_K\right)\ }$ or }\fi $\frac{\bar{\gamma}_K+1}{1+\beta \bar{\gamma}_K}e^{\frac{\bar{\gamma}_K+1}{1+\beta \bar{\gamma}_K}}=e\left(\bar{\gamma}_K+1\right),$
the solution to which is
\begin{equation*}
\beta =\frac{\bar{\gamma}_K+1}{\bar{\gamma }_K W_0\left(e\left(\bar{\gamma }_K+1\right)\right)}-\frac{1}{\bar{\gamma}_K},
\end{equation*}
where the principal branch of the Lambert-W function is used because its argument is always positive.

Now, imposing (\myref[eqBetaConstraintEq]) upon this equation gives (\myref[eqOptimalBeta]).
It can be shown that for $\gamma_K>0$ we have $\beta>0$.
\textblue{On the other hand, we earlier assumed $P_b \le Bs_\mathrm{max}$.}
Therefore, (\myref[UeqBetaLimitEq]) is satisfied as well.

\hfill $\blacksquare$

In a power-constrained system, the first argument of the min operator and in a BW-constrained system the second argument takes hold.

\subsection{CSI Feedback Phase Time Length Ratio}
Increasing $\alpha$ improves CSI knowledge at the HAP which increases the WIT data rate.
At the same time, however, it decreases the WIT phase time length which lowers the WIT data rate.
So, an optimal value for $\alpha$ exists which we will subsequently find.

Using (\myref[eqOptimizationTwo]), we derive an iterative solution for the optimal CSI feedback phase time length ratio in theorem \myref[thm:optimal.alpha].
Nevertheless, before that, we shall prove the optimization problem (\myref[eqOptimizationTwo]) is convex w.r.t $\alpha$\footnote{Note that in lemma \myref[lem:convexity1] we showed that the problem is quasiconvex.
But quasi-convexity is not enough for the convergence of an iterative solution to a global optimum.}.

\begin{lem}
\label{lem:Convexity2}
Optimization problem (\myref[eqOptimizationTwo]) is convex w.r.t $\alpha$.
\end{lem}

\emph{Proof}: Please refer to Appendix \myref[app.convexity.2].
\hfill $\blacksquare$

Now we can proceed to find the optimal $\alpha$.

\def \eqOptimalAlphaOne{
\begin{equation}
\alpha^* = \underset{\alpha \in \{\alpha_1,\alpha_{\infty}\}}{\arg\max}\, r_{w,K},
\label{eqOptimalAlphaOne}
\end{equation}
}
\def \eqOptimalAlphaTwo{
\begin{subequations}
\label{eqOptimalAlphaTwo}
\begin{align}
\alpha_{\infty} &= \lim_{n\to\infty} \alpha_n,
\label{eqOptimalAlphaTwo.infty}\\
\alpha_{n+1} &=\frac{\frac{M-1}{TB(1-\beta)}\log_2\left(\frac{\gamma_K^\mathrm{maxloss}\left(TB\ln2 \frac{1-\beta}{M-1}+1\right)}{1+{\gamma}_K^{\mathrm{max}}}\right)}{\log_2\left(1+{\gamma}_K^{\mathrm{max}}-\frac{(1+{\gamma}_K^{\mathrm{max}}){\gamma}_K^{\mathrm{maxloss}}}{(1+{\gamma}_K^{\mathrm{max}})^{\alpha_n\left(\frac{TB}{M-1}\right)+1}-\alpha_n\left(\frac{TB}{M-1}\right){\gamma}_K^{\mathrm{maxloss}}}\right)},
\label{eqOptimalAlphaTwo.n}\\
\alpha_1 &= 0,
\label{eqOptimalAlphaTwo.1}
\end{align}
\end{subequations}
}
\begin{thm}
\label{thm:optimal.alpha}
The optimal $\alpha$, denoted by $\alpha^*$ is given by
\eqOptimalAlphaOne
where
\eqOptimalAlphaTwo
and $n\in\mathbb{N}$ is the iteration number.
Note that, in practice, a few iterations are sufficient.
\end{thm}

\emph{Proof}:
Rewriting the $K$-$th$ element of (\myref[eqFundamentalOne]) in terms of $r_{w,K}$ gives
\begin{equation*}
r_{w,K}=\left(1-\alpha \right)\left(1-\beta \right)B{\log_2 \left(1+\gamma^{\mathrm{max}}_K-\gamma^{\mathrm{maxloss}}_K2^{-\frac{\alpha }{1-\alpha }\left(\frac{Tr_{w,K}}{M-1}\right)}\right)\ }.
\end{equation*}
Taking derivative with respect to $\alpha$ 
\ext{
results in
\begin{equation*}
\frac{\partial r_{w,K}}{\partial \alpha }=\frac{\left(1-\alpha \right)\left(1-\beta \right)B}{\ln 2}\frac{-\gamma^\mathrm{maxloss}_k\frac{\partial \left(2^{-\frac{\alpha }{1-\alpha }\left(\frac{Tr_{w,K}}{M-1}\right)}\right)}{\partial \alpha }}{1+\gamma^\mathrm{max}_K-\gamma^{\mathrm{maxloss}}_K2^{-\frac{\alpha }{1-\alpha }\left(\frac{Tr_{w,K}}{M-1}\right)}}-r_K
\end{equation*}
But
\begin{multline*}
\frac{\partial \left(2^{-\frac{\alpha }{1-\alpha }\left(\frac{Tr_{w,k}}{M-1}\right)}\right)}{\partial \alpha }=-\frac{\ln 2\ 2^{-\frac{\alpha }{1-\alpha }\left(\frac{Tr_{w,k}}{M-1}\right)}}{M-1}\times\\
\left(\frac{1}{{\left(1-\alpha \right)}^2}Tr_{w,k}+\frac{\alpha }{1-\alpha }T\frac{\partial r_{w,k}}{\partial \alpha }\right)
\end{multline*}
Substituting
\begin{multline*}
\frac{\partial r_{w,K}}{\partial \alpha }=B\left(1-\alpha \right)\left(1-\beta \right)\times\\
\frac{\gamma^{\mathrm{maxloss}}_K \frac{2^{-\frac{\alpha }{1-\alpha }\left(\frac{Tr_{w,K}}{M-1}\right)}}{M-1}\left(\frac{1}{{\left(1-\alpha \right)}^2}Tr_{w,K}+\frac{\alpha }{1-\alpha }T\frac{\partial r_{w,K}}{\partial \alpha }\right)}{1+\gamma^{\mathrm{max}}_K-\gamma^{\mathrm{maxloss}}_K2^{-\frac{\alpha }{1-\alpha }\left(\frac{Tr_{w,K}}{M-1}\right)}}-r_K
\end{multline*}
}{}
and setting $\frac{\partial r_{w,K}}{\partial \alpha }=0$ results in
\begin{equation*}
\frac{TB{\gamma }^{\mathrm{maxloss}}_K\frac{2^{-\frac{\alpha }{1-\alpha }\left(\frac{Tr_{w,K}}{M-1}\right)}}{M-1}\left(\frac{1-\beta}{1-\alpha}r_{w,K}\right)}{1+\gamma^{\mathrm{max}}_K-\gamma^{\mathrm{maxloss}}_K2^{-\frac{\alpha }{1-\alpha }\left(\frac{Tr_{w,K}}{M-1}\right)}} = r_K.
\end{equation*}
\ext{
But$\ \frac{1}{\left(1-\alpha \right)}r_{w,K}=r_K$, which means
\begin{equation*}
BT\left(1-\beta \right)\gamma^{\mathrm{maxloss}}_K\frac{2^{-\alpha \left(\frac{Tr_K}{M-1}\right)}}{M-1}=1+{\gamma }^{\mathrm{max}}_K-\gamma^{\mathrm{maxloss}}_K2^{-\alpha \left(\frac{Tr_K}{M-1}\right)}
\end{equation*}
Moving the exponentials to one side and dividing by the common factor gives
\begin{equation*}
2^{-\alpha \left(\frac{Tr_K}{M-1}\right)} = \frac{1+\gamma^\mathrm{max}_K}{\gamma^{\mathrm{maxloss}}_K\left(TB\frac{1-\beta}{M-1}+1\right)}
\end{equation*}
taking the logarithm
\begin{equation*}
\alpha =-\frac{M-1}{Tr_K}{\log_2 \left(\frac{1+\gamma^\mathrm{max}_K}{\gamma^{\mathrm{maxloss}}_K\left(TB\frac{1-\beta}{M-1}+1\right)}\right)\ }
\end{equation*}
We can now substitute for $r_K$ using (\myref[eqDataRateWITThree])
}{}
Using (\myref[eqDataRateWITThree]) as well as the definition for $r_{w,K}$, gives
\begin{equation*}
\alpha_\infty=\frac{\frac{M-1}{TB(1-\beta)}{\log_2 \left(\frac{\gamma^\mathrm{maxloss}_K\left(TB\frac{1-\beta}{M-1}+1\right)}{1+\gamma^{\mathrm{max}}_K}\right)\ }}{\log_2 \left(1+\gamma^\mathrm{max}_K-\frac{\left(1+\gamma^{\mathrm{max}}_K\right)\gamma^\mathrm{maxloss}_K}{{\left(1+\gamma^{\mathrm{max}}_K\right)}^{\alpha_\infty\left(\frac{TB}{M-1}\right)+1}-\alpha_\infty\left(\frac{TB}{M-1}\right)\gamma^{\mathrm{maxloss}}_K}\right)\ },
\end{equation*}
%
On the other hand, $r_{w,K}$ is 0 for $\alpha = 1$ and so we do not have to check for this boundary value.
\hfill $\blacksquare$

\subsection{Algorithm}
\textblue{
The formulae for the optimum value of $\alpha$, $\beta$, and $\boldsymbol{\xiup}$ depend on $(\alpha, \beta, \boldsymbol{\xiup})$, $(\boldsymbol{\xiup})$, and $(\alpha, \beta, \boldsymbol{\xiup})$ respectively.
Thus, a new set of values computed for $\alpha$ and $\beta$ makes the value computed for $\boldsymbol{\xiup}$ no longer optimum and vice versa.
As a result, the optimal parameters should be found iteratively.
}
arx{Here we present an algorithm that will calculate the optimum parameters $\boldsymbol{\xiup}^*$, $\alpha^*$, and $\beta^*$.
The procedure is outlined in Alg. 1.}
\arx{\textblue{Note that the parameters have been initialized with their asymptotic values to be calculated in the next section.
The exception is $\beta$ where an initial value of zero makes the data rate fall to zero, rendering the feedback useless.}}{Note that it is best to initialize the parameters with their asymptotic values to be calculated in the next section.}

\arx{
\begin{algorithm}
\label{algorithm}
\DontPrintSemicolon
\KwData{$B, T, M, s_{\mathrm{max}}, \boldsymbol{d}, \mathcal{K}$}
\KwResult{$\boldsymbol{\xiup}^*$, $\alpha^*$, $\beta^*$}

\Begin{
Initialize $\epsilon_1, \epsilon_2$\tcp*[f]{Error Tolerances}\;
Initialize $c_0, d_0, \delta$\tcp*[f]{\textblue{Constants}}\;
Compute $\boldsymbol{b}$ from (\myref[eqLongTermFading])\;
$k\leftarrow0$\;
$\alpha_{\mathrm{new}}^{(k)} \leftarrow 0$\;
$\beta^{(k)} \leftarrow \frac{P_b}{Bs_{\mathrm{max}}}$\;
$\boldsymbol{\xiup}^{(k)} \leftarrow \| \boldsymbol{b}^{-2} \|_1^{-1} \boldsymbol{b}^{-2}$\;
\While{$\left\| \boldsymbol{\xiup}^{(k)}-\boldsymbol{\xiup}^{(k+1)} \right\|_2 < \epsilon_1$}{
$k \leftarrow k + 1$\;
Compute $\beta^{(k)}$ from (\myref[eqOptimalBeta])\;

\While{$|\alpha_{\mathrm{new}}^{(k)} -\alpha_{\mathrm{old}}^{(k)} | < \epsilon_2$}{
$\alpha_\mathrm{old}^{(k)} \leftarrow \alpha_\mathrm{new}^{(k)}$\;
Compute $\alpha_{\mathrm{new}}^{(k)}$ from (\myref[eqOptimalAlphaOne])\;
}

$\boldsymbol{\mathcal{K}}_{\mathrm{new}}^f \leftarrow \left\{1,...,M\right\}$\;
$\boldsymbol{\mathcal{K}}_{\mathrm{new}}^u \leftarrow \varnothing$\;

$\boldsymbol{\mathcal{K}}_{\mathrm{old}}^f \leftarrow \varnothing$\;
$\boldsymbol{\mathcal{K}}_{\mathrm{old}}^u \leftarrow \left\{1,...,M\right\}$\;

\While{$\mathcal{K}_{\mathrm{old}}^f=\mathcal{K}_{\mathrm{new}}^f$}{
$\boldsymbol{\xiup}_{\boldsymbol{\mathcal{K}}^f} \leftarrow \boldsymbol{M}^{-1}_{\boldsymbol{\mathcal{K}}^f,\boldsymbol{\mathcal{K}}^f}\boldsymbol{b}^{-2}_{\boldsymbol{\mathcal{K}}^f}$\;
$\boldsymbol{\xiup}_{\boldsymbol{\mathcal{K}}^f} \leftarrow \left\|\boldsymbol{\xiup}_{\boldsymbol{\mathcal{K}}^f}\right\|_1^{-1}\boldsymbol{\xiup}_{\boldsymbol{\mathcal{K}}^f}$ \tcp*[f]{Normalizing the vector}\;
$\boldsymbol{\xiup}_{\boldsymbol{\mathcal{K}}^u} \leftarrow \boldsymbol{0}$\;

$\boldsymbol{\mathcal{K}}_{\mathrm{old}}^f \leftarrow \boldsymbol{\mathcal{K}}_{\mathrm{new}}^u$\;
$\boldsymbol{\mathcal{K}}_{\mathrm{old}}^f \leftarrow \boldsymbol{\mathcal{K}}_{\mathrm{new}}^u$\;

$\boldsymbol{\mathcal{K}}_{\mathrm{new}}^f \leftarrow (f_1,...,f_k) \subset \boldsymbol{\mathcal{K}} \mathrm{~where~} \boldsymbol{\xiup}_{f_i} > \boldsymbol{0}$\;
$\boldsymbol{\mathcal{K}}_{\mathrm{new}}^u \leftarrow (f_1,...,f_l) \subset \boldsymbol{\mathcal{K}} \mathrm{~where~} \boldsymbol{\xiup}_{u_i} \le \boldsymbol{0}$\;

}
}
}
$\boldsymbol{r}_{w}=(1-\alpha^{(k)}_\mathrm{new})\boldsymbol{r}$\;
\caption{MU-MISO WPCN optimization}
\end{algorithm}
}{}

\section{Asymptotic Behavior}
In this section we will study the asymptotic behavior of our system as the number of HAP antennas $M$ goes to infinity.
This is especially important because today's trend toward multiple antenna communication system's design is to increase the number of antennas to gain the many benefits of operating in the massive multiple-input-multiple-output (MIMO) regime \cite{06375940}.
\arx{\textblue{Additionally, as mentioned previously, we use these asymptotics for initialization of the optimization algorithm.}}{}

\textblue{
In corollary \myref[cor:asymptotic.xiup], we obtain the asymptotic value of the energy allocation weight vector.
}

\def \eqAsymptoticXiup{
\begin{equation}
\label{eqAsymptoticXiup}
\boldsymbol{\xiup}^\mathrm{asym} = {\left\|{\boldsymbol{b}}^{-2}\right\|}_1^{-1}{\boldsymbol{b}}^{-2}.
\end{equation}
}

\begin{cor}
\label{cor:asymptotic.xiup}
As $M$ tends to infinity, the asymptotic value of the optimal $\boldsymbol{\xiup}$ is given by 
\eqAsymptoticXiup
\end{cor}

\emph{Proof}:
Assuming $M$ goes to infinity,
we can show that
$$\alpha \left(\frac{TB}{M-1}\right)\gamma^{\mathrm{maxloss}}_k \rightarrow c_1 \alpha M \rightarrow \infty, $$
where $c_1$ is a constant.
But, according to (\myref[eqFeedbackErrorThree]),
$\boldsymbol{\sigma}_{u,f}\rightarrow \boldsymbol{0}$,
and as a result
$\boldsymbol{M} \rightarrow M \boldsymbol{I}$.
Substituting this into (\myref[eqOptimalXiupOne]), (\myref[eqAsymptoticXiup]) results.
\hfill $\blacksquare$

Note that, in contrast to $\boldsymbol{\xiup}^*$, non of the elements of $\boldsymbol{\xiup}^\mathrm{asym}$ ever become zero.
This is due to the fact that with an infinite number of HAP antennas, the interference power goes to zero.
This effect is alternatively explained by the asymptotic behavior of the fairness radius described in the following corollary which we state without proof.

\def \eqAsymptoticFairnessRadius{
\begin{equation}
\label{eqAsymptoticFairnessRadius}
r_f^\mathrm{asym} = O\left(M^{-\frac{1}{2\delta}}\right) \rightarrow 0.
\end{equation}
}
\begin{cor}
\label{cor:asymptotic.r.f}
As $M$ goes to infinity, the fairness radius $r_f$ becomes
\eqAsymptoticFairnessRadius
\end{cor}

\emph{Proof:}
The proof is simple and follows by taking the limit of (\myref[eqFairnessRadiusOne]) as $M$ tends to infinity.
\hfill $\blacksquare$

From (\myref[eqEnergyVectorTwo]), for small $M$, the interference energy is not negligible compared with the beamformed energy which means sufficient interference energy to supply the WDs can reach up to a large distance from the HAP.
As a result, the fairness radius is large.
However, as the number of HAP antennas $M$ is increased, the interference power decreases, making it sufficient to supply the power of WDs only at small distances from the HAP.
Therefore, the fairness radius decreases.
As $M$ goes to infinity, the interference energy vanishes, making the fairness radius reach zero.

Next, the asymptotic value of $\beta$ is given.

\def \eqAsymptoticBeta{
\begin{equation}
\beta^\mathrm{asym} \rightarrow \frac{1}{W_0\left(e\bar{\gamma }_K\right)} \rightarrow \frac{1}{\ln\left(\bar{\gamma }_K\right)} \rightarrow 0.
\label{eqAsymptoticBeta}
\end{equation}
}
\begin{cor}
\label{cor:asymptotic.beta}
As $M$ goes to infinity, the optimal DL BW ratio $\beta$ tends to
\eqAsymptoticBeta
\end{cor}

\emph{Proof}:
As $M$ increases without bound,
%
%
the first argument of the min operator in (\myref[eqOptimalBeta]) takes hold and goes to zero.
Its limit is given by (\myref[eqAsymptoticBeta])
%
%
\hfill $\blacksquare$

This means that as the number of HAP antennas increases unboundedly, the HAP, instead of modulated power signals, may transfer the power using a single tone.


In the following corollary, we find the asymptotic behavior of the optimal CSI feedback phase time length ratio.

\def \eqAsymptoticAlpha{
\begin{equation}
\label{eqAsymptoticAlpha}
\alpha^\mathrm{asym} \rightarrow \frac{\ln{2}}{W_0\left(\frac{TB\gamma_K^{\mathrm{max}}\ln{(\gamma_K^{\mathrm{max}})}\ln{2}}{M-1}\right)} \rightarrow \frac{\ln{2}}{\ln\left(M\right)} \rightarrow 0.
\end{equation}
}
\begin{cor}
\label{cor:AsymptoticAlpha}
As $M$ goes to infinity, the asymptotic value of the optimal $\alpha$ is described by 
\eqAsymptoticAlpha
\end{cor}

\textblue{\emph{Proof}: The proof is easy and follows by taking the limit of $\alpha$}.
\hfill $\blacksquare$

\section{Simulation Results}
In what follows, we present the results for three different simulation scenarios, showing the accuracy of the solution of the forward problem in the first, the accuracy of the solution of the optimization problem in the second, and comparing variations of the optimal WIT data rates for all WDs versus the number of antennas in the third.
Unless otherwise stated, we set the following parameters
$M=10$,
$K = 4$, 
$P_b = 10 \mathrm{Watt}$, 
$T = 1 \mathrm{ms}$, 
$\sigma^2 = -120 \mathrm{dBm}$,
$s_{\mathrm{max}}=10^{-4}\mathrm{W/Hz}$,
 and $B=100 \mathrm{KHz}$.
\arx{Note that with these parameters, $P_b$ is not larger than \textblue{$Bs_{\mathrm{max}}=10 \mathrm{W}$}.}{}As for the long-term fading model (\myref[eqLongTermFading]), we assume $c_0 = 10^{-3}$, $\delta = 3$ and the distances of the HAP to the $WD_k, ~ \forall k\in \boldsymbol{\mathcal{K}}=\{1,2,3,4\}$ is $d_k = 4m + (k-1) \times 2m$. 

\subsection{Forward Problem}

\begin{table}[t]
\centering
\caption{\textblue{WIT data rates of all WDs in different sub-scenarios for simulation versus theory.}}
\label{table:simulation}
\ifdouble
\resizebox{250pt}{!}{\begin{tabular}{ l c c c c c c }
\else
\resizebox{300pt}{!}{\begin{tabular}{ l c c c c c c }
\fi
 \hline
 Method & Sub-scenario & $r_{w,1}$ & $r_{w,2}$ & $r_{w,3}$ & $r_{w,4}$\\ 
 \hline
 \hline
 Simulation 							  & 1 & 1.1740  &  0.5309  &  0.3669  &  0.2036  \\ 
 Analytic (\myref[eqDataRateWITThree])    & 1 & 1.1826  &  0.6000  &  0.3914  &  0.2400  \\ 
  \hline
 Simulation                               & 2 & 0.8501  &  0.8757  &  0.3257  &  0.1905  \\ 
 Analytic (\myref[eqDataRateWITThree])    & 2 & 0.8992  &  0.8808  &  0.3914  &  0.2400  \\ 
  \hline
 Simulation  						 	  & 3 & 0.8342  &  0.5297  &  0.6586  &  0.2001  \\ 
 Analytic (\myref[eqDataRateWITThree])    & 3 & 0.8992  &  0.6000  &  0.6644  &  0.2400  \\ 
  \hline
 Simulation  							  & 4 & 0.8308  &  0.5308  &  0.3395  &  0.4577  \\ 
 Analytic (\myref[eqDataRateWITThree])    & 4 & 0.8992  &  0.6000  &  0.3914  &  0.4932  \\ 
 \hline
 Simulation  							  & 5 & 1.0369  &  0.7207  &  0.4922  &  0.2698  \\ 
 Analytic (\myref[eqDataRateWITThree])    & 5 & 1.0437  &  0.7414  &  0.5240  &  0.3528  \\ 
 \hline
\end{tabular}}

\end{table}

Here, we set the decision variables and measure the resulting WIT data rates.
The DL BW ratio is set to $\beta=0.1$, the CSI feedback phase time length ratio is set to $\alpha = 0.05$, and \textblue{the energy allocation weight vector is set to $\boldsymbol{\xiup}_k=\boldsymbol{e}_k,~ k \in \boldsymbol{\mathcal{K}}$,  $\boldsymbol{\xiup}_5=4^{-1}\boldsymbol{1}_4$ in five different sub-scenarios.}
Note that these parameters satisfy the required constraints (\myref[eqXiPositivityConstraintEq])-(\myref[eqBetaConstraintEq]).

\textblue{
In order to verify the results, 1000 channel realizations are used for each case.
The simulation, and the analytic results based on (\myref[eqDataRateWITThree]) are illustrated in table \myref[table:simulation].
As can be seen, the analytic results are in decent  agreement with the simulation results.
}

\subsection{Optimization Problem}

\begin{figure*}[t]
\centering
\ifdouble
  	\subfigure[]{\includegraphics[scale=0.55]{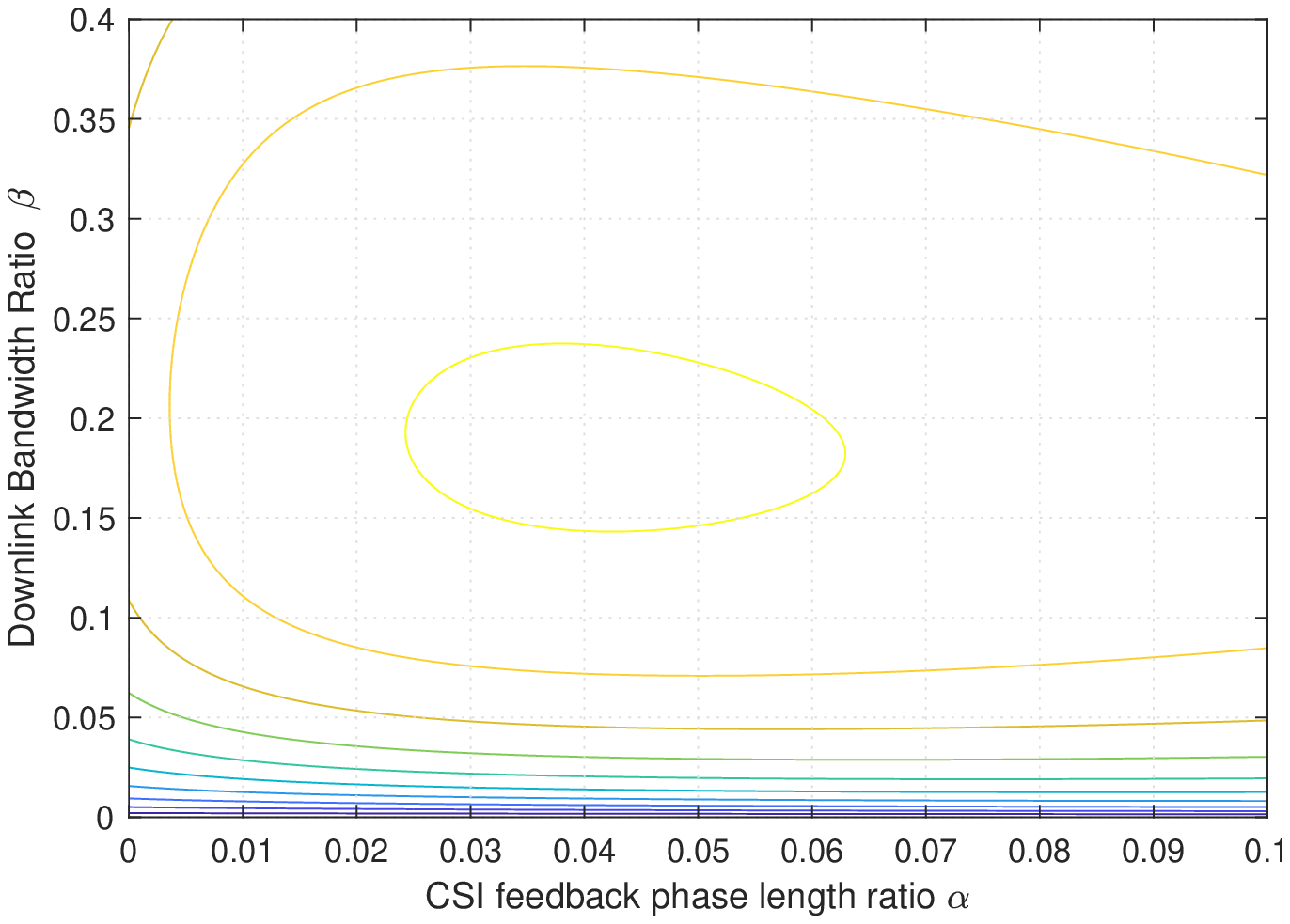}}
  	\quad
  	\subfigure[]{\includegraphics[scale=0.55]{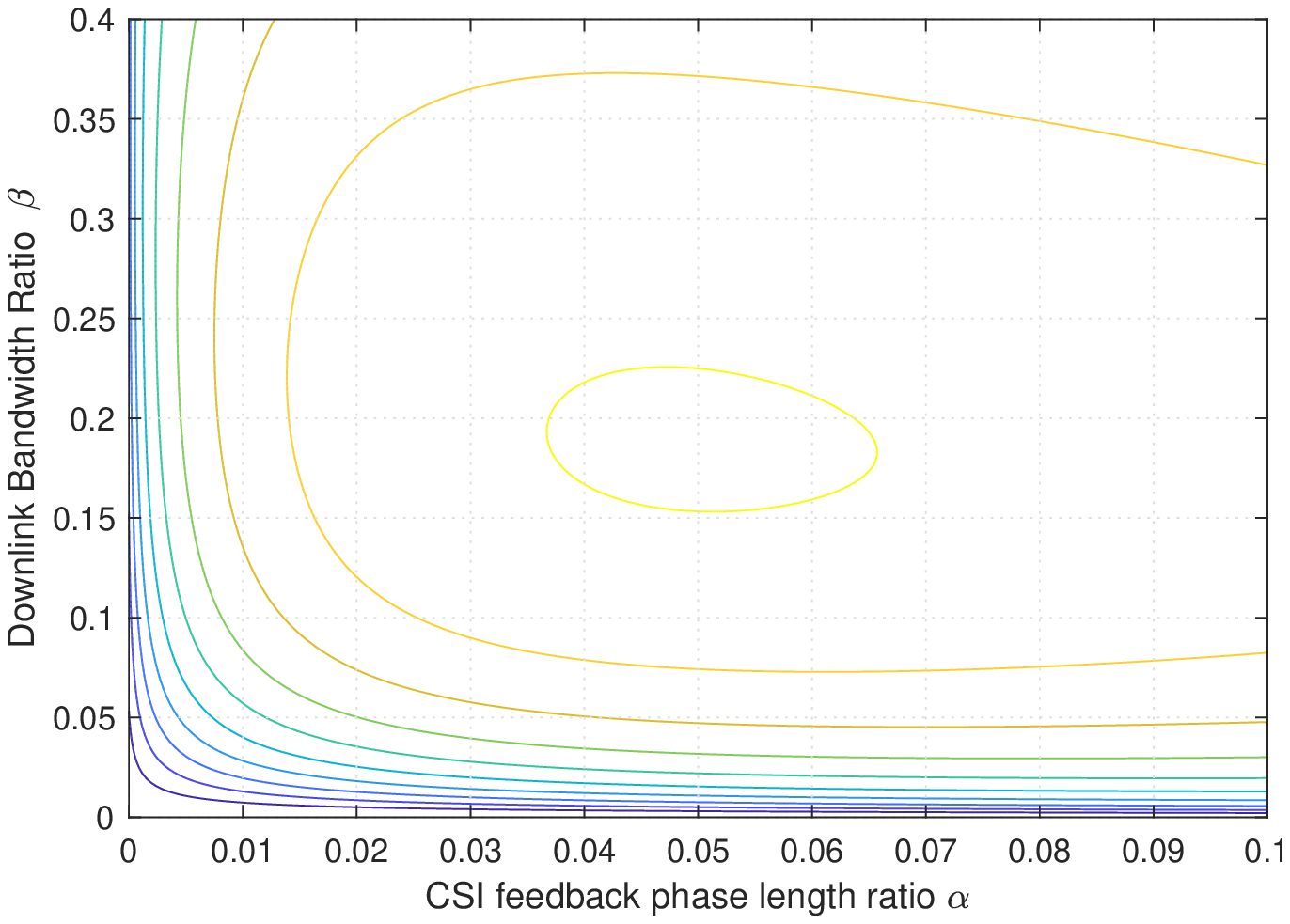}}
\else
  	\subfigure[]{\includegraphics[trim=50 0 20 0, scale=0.55]{sim2A.eps}}
  	\quad
  	\subfigure[]{\includegraphics[trim=50 0 20 0, scale=0.55]{sim2B.eps}}
\fi
\caption{Variations of WIT data rates versus CSI feedback phase time length ratio $\alpha$ and DL BW ratio $\beta$ for (a) wireless device 1 and (b) wireless device 2}
\label{fig.sim1}
\end{figure*}

\textblue{
Next, we compare the UL WIT data rate versus $\alpha$ and $\beta$ for different WDs obtained from theory. 
The fairness radius is 6.03 which means $\boldsymbol{\mathcal{K}}^u = \{1,2\}$ and $\boldsymbol{\mathcal{K}}^f = \{3,4\}$.
Therefore, the data rates for $WD_3$ and $WD_4$ should agree reasonably well around the optimal CSI feedback phase time length ratio $\alpha^*$ and the optimal UL BW ratio $\beta^*$ while $WD_1$ and $WD_2$ should have higher WIT data rates. 
For this reason, only the WIT data rates of $WD_1$ and $WD_2$ are plotted in Fig. \myref[fig.sim1].
The optimal $\alpha$ and $\beta$ calculated from (\myref[eqOptimalAlphaOne]) and (\myref[eqOptimalBeta]) are obtained as $0.0558$ and $0.1802$ whereas those calculated via numerical search are $0.0490$ and $0.1874$ respectively.
Finally, the minimum feedback bit rate among the WDs at the optimal point is about 25 bits.
}




\subsection{Optimal WIT Data Rates versus $M$}

\begin{figure}[t]
\ifdouble{
\includegraphics[scale=0.6]{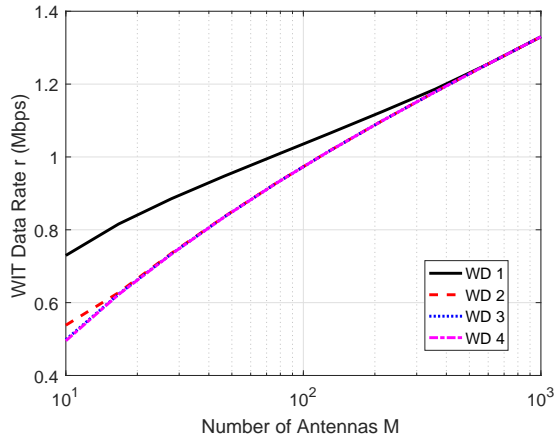}
}
\else{
\includegraphics[scale=0.55]{sim3.eps}
}
\fi
\centering
\caption{WIT rate fairness for wireless devices 1 to 4 versus the number of antennas M}
\label{fig.sim3}
\end{figure}

Finally, this subsection examines the optimal WIT data rates for all WDs versus the number of antennas $M$.
As can be seen in Fig. \myref[fig.sim3], initially, the WIT data rates are relatively different.
Specifically, $r_{w,4} \approx r_{w,3}<r_{w,2}<r_{w,1}$ which is due to the fact that $WD_3$ and $WD_4$ are in the fair region but $WD_1$ and $WD_2$ are not.
But, as the number of antennas increases, the fairness radius decreases, hence reducing the WIT data rate gaps until eventually at $M\approx475$, they reach zero and do not change afterward.

\section{Conclusion}
In this paper a MU-MISO WPC network operating in FDD mode was studied.
The proposed scheme optimized the energy allocation weight vector, the amount of CSI feedback phase time length ratio, and the UL BW ratio needed to maximize the minimum UL throughput.
Based on the level of fairness achieved, we divided the area around the HAP into two regions: a fair region, and an unfair region.
It was shown that by increasing the number of HAP antennas, better fairness is accomplished.
In addition, the amount of feedback and the UL BW ratio reduce asymptotically with the number of HAP antennas.

\ifCLASSOPTIONcaptionsoff
  \newpage
\fi

\appendices
\section{Proof for Lemma \myref[lem:pareto.optimal.beamformer] Pareto Optimal Beamformer}
\label{app.pareto.optimal.beamformer}
Suppose the beamformer is not pareto optimal. Then, it can be written in the most general form as
$\boldsymbol{w}=\tilde{\boldsymbol{G}}_d{\boldsymbol{\xiup}^{\prime}}^{1/2}+\boldsymbol{U}{\boldsymbol{\xiup}^{\prime\prime}}^{1/2},$
where $\boldsymbol{0}_k \le \boldsymbol{\xiup}^\prime$ and $\boldsymbol{0}_k \le \boldsymbol{\xiup^{\prime\prime} }$ are arbitrary vectors satisfying $\|\boldsymbol{\xiup }^{\prime}\|_1+\|\boldsymbol{\xiup^{\prime\prime} }\|_1=1$; and $\boldsymbol{U}\in \mathbb{R}^{M\times(M-K)} $ is a matrix for which we have $\boldsymbol{U}^T \boldsymbol{\tilde{G}}_d = \boldsymbol{0}\boldsymbol{0}^T$ and $\boldsymbol{U}^T \boldsymbol{U} = \boldsymbol{I}$.

Following a similar procedure to that in \cite[lemma 1]{07009979} and using
(\myref[eqChannelOuterProd]) and (\myref[eqEnergyScalar]) we can calculate the expected harvested energy for this general beamformer
\begin{equation*}
\boldsymbol{\varepsilon}^\prime=T \beta s_{\mathrm{max}} \boldsymbol{b}  \odot (\boldsymbol{M} \boldsymbol{\xiup}^\prime )+T\beta s_{\mathrm{max}} \|\boldsymbol{\xiup^{\prime\prime}}\|_1 \boldsymbol{b} \odot \boldsymbol{\sigma }^2_{u,f},
\end{equation*}
where the mixing power matrix $\boldsymbol{M}$ is defined in lemma \myref[lem:EnergyVector].
Now, let us consider a beamformer with the proposed pareto optimal structure in (\myref[eqParetoOptimalBeamformer]) and 
$\boldsymbol{\xiup} = {\|{\boldsymbol{\xiup}^{\prime}}\|_1}^{-1}\boldsymbol{\xiup }^{\prime}.$
The amount of energy this beamformer can deliver to WDs is given by
\begin{equation*}
\boldsymbol{\varepsilon} = \frac{T \beta s_{\mathrm{max}}}{\|{\boldsymbol{\xiup}^{\prime}}\|_1} \boldsymbol{b} \odot(\boldsymbol{M}\boldsymbol{\xiup}^\prime ).
\end{equation*}
We can calculate the difference of the harvested energy by these two beamformers
\begin{equation*}
\boldsymbol{\varepsilon }-\boldsymbol{\varepsilon }^{\prime}=T \beta s_{\mathrm{max}} {\|{\boldsymbol{\xiup^{\prime\prime}}}\|_1} \boldsymbol{b} \odot\left({\|{\boldsymbol{\xiup}^{\prime}}\|_1}^{-1} \boldsymbol{M}\boldsymbol{\xiup}^\prime-\boldsymbol{\sigma }^2_{u,f}\right).
\end{equation*}

In order for the elements of this vector to be positive, we need to have
\begin{equation*}
{\sigma }^2_{u,f,k}\le \frac{M{\xiup}^{\prime}_k\left(1-{\sigma }^2_{u,f,k}\right)+\|\boldsymbol{\xiup}^\prime\|_1-\xiup^\prime_k}{\|{\boldsymbol{\xiup}^{\prime}}\|_1}.
\end{equation*}
A sufficient condition would be
${\sigma }^2_{u,f,k}\le \frac{\left(M-1\right){\xiup}^{\prime}_k+1}{M{\xiup}^{\prime}_k+1},$
which, using (\myref[eqFeedbackErrorTwo]) gives
\begin{equation*}
-\frac{n_k}{M-1}\le {\log_2 \left(\frac{\left(M-1\right){\xiup}^{\prime}_k+1}{M{\xiup}^\prime_k+1}\right)\ },
\end{equation*}
for which a sufficient condition is
\begin{equation*}
n_k \ge \frac{1}{\ln 2} \cong 0.7.
\end{equation*}

\section{Proof for Lemma \myref[lem:UL.data.rate] Uplink Data Rate}
\label{app.UL.data.rate}

\label{app:UL.data.rate}

Exponentiating both sides of (\myref[eqFundamentalOne]) and substituting for $r_k^{\mathrm{max}}$ results in
\begin{equation*}
2^{\left(\frac{r_k}{B}\right)}=1+\gamma^{\mathrm{max}}_k-\gamma^{\mathrm{maxloss}}_k2^{-\alpha \left(\frac{Tr_k}{M-1}\right)}.
\end{equation*}
Defining $x = 2^{\left(\frac{r_k}{B}\right)}$ and assuming it is greater than 1, it can be iteratively approximated as follows
\begin{align*}
x_1&=1+\gamma^{\mathrm{max}}_k,\\
x_n&=1+\gamma^{\mathrm{max}}_k-\gamma^{\mathrm{maxloss}}_k x_{n-1}^{-\alpha \left(\frac{TB}{M-1}\right)},
\end{align*}
where $n\in\mathbb{N}$, is the iteration number.
We can further linearize the expressions for $x_{n+1}$.
To see how this is possible, consider
\begin{equation*}
x_3=1+\gamma^\mathrm{max}_k-\gamma^\mathrm{maxloss}_k{\left(1+\gamma^\mathrm{max}_k-\gamma^\mathrm{maxloss}_k{x_1}^{-\alpha \left(\frac{TB}{M-1}\right)}\right)}^{-\alpha \left(\frac{TB}{M-1}\right)}.
\end{equation*}
Substituting for $x_1$, factoring out $\left(1+\gamma^\mathrm{max}_k\right)^{-\alpha \left(\frac{TB}{M-1}\right)}$, and assuming $\gamma^{\mathrm{maxloss}}_k{\left(1+\gamma^{\mathrm{max}}_k\right)}^{-\alpha \left(\frac{TB}{M-1}\right)-1}\ll 1$, we can derive a first-order approximation for $x_3$ as follows
\ifdouble
\begin{multline*}
x_3\cong 1+\gamma^\mathrm{max}_k-\gamma^{\mathrm{maxloss}}_k{\left(1+\gamma^{\mathrm{max}}_k\right)}^{-\alpha \left(\frac{TB}{M-1}\right)}\\
\left(1+\alpha \left(\frac{TB}{M-1}\right)\gamma^{\mathrm{maxloss}}_k{\left(1+{\gamma }^{\mathrm{max}}_k\right)}^{-\alpha \left(\frac{TB}{M-1}\right)-1}\right),
\end{multline*}
\else
\begin{equation*}
x_3\cong 1+\gamma^\mathrm{max}_k-\gamma^{\mathrm{maxloss}}_k{\left(1+\gamma^{\mathrm{max}}_k\right)}^{-\alpha \left(\frac{TB}{M-1}\right)}
\left(1+\alpha \left(\frac{TB}{M-1}\right)\gamma^{\mathrm{maxloss}}_k{\left(1+{\gamma }^{\mathrm{max}}_k\right)}^{-\alpha \left(\frac{TB}{M-1}\right)-1}\right),
\end{equation*}
\fi
which can be rewritten as
\ifdouble
\begin{equation*}
\begin{split}
x_3&\cong 1+\gamma^{\mathrm{max}}_k-\gamma^{\mathrm{maxloss}}_k{\left(1+\gamma^{\mathrm{max}}_k\right)}^{-\alpha \left(\frac{TB}{M-1}\right)}\\
&-\alpha \left(\frac{TB}{M-1}\right){\left(\gamma^{\mathrm{maxloss}}_k\right)}^2{\left(1+\gamma^{\mathrm{max}}_k\right)}^{-2\alpha \left(\frac{TB}{M-1}\right)-1}.
\end{split}
\end{equation*}
\else
\begin{equation*}
x_3\cong 1+\gamma^{\mathrm{max}}_k-\gamma^{\mathrm{maxloss}}_k{\left(1+\gamma^{\mathrm{max}}_k\right)}^{-\alpha \left(\frac{TB}{M-1}\right)}
-\alpha \left(\frac{TB}{M-1}\right){\left(\gamma^{\mathrm{maxloss}}_k\right)}^2{\left(1+\gamma^{\mathrm{max}}_k\right)}^{-2\alpha \left(\frac{TB}{M-1}\right)-1}.
\end{equation*}
\fi
Doing this operation $n-1$ times for $x_n$ results in
\begin{equation*}
x_n \cong 1+\gamma^{\mathrm{max}}_k-\frac{1+\gamma^{\mathrm{max}}_k}{\alpha \left(\frac{TB}{M-1}\right)}\sum^n_{i=1}{{\left(\frac{\alpha \left(\frac{TB}{M-1}\right)\gamma^{\mathrm{maxloss}}_k}{{\left(1+\gamma^{\mathrm{max}}_k\right)}^{1+\alpha \left(\frac{TB}{M-1}\right)}}\right)}^i}.
\end{equation*}
Letting $n\rightarrow \infty$ gives
\begin{equation*}
x \cong 1+\gamma^{\mathrm{max}}_k-\frac{\left(1+\gamma^{\mathrm{max}}_k\right)\gamma^{\mathrm{maxloss}}_k}{{\left(1+\gamma^{\mathrm{max}}_k\right)}^{1+\alpha \left(\frac{TB}{M-1}\right)}-\alpha \left(\frac{TB}{M-1}\right)\gamma^{\mathrm{maxloss}}_k}.
\end{equation*}
As a result,
\begin{equation*}
{\sigma}_{u,f,k}^2 = \frac{1+\gamma^{\mathrm{max}}_k}{{\left(1+\gamma^{\mathrm{max}}_k\right)}^{1+\alpha \left(\frac{TB}{M-1}\right)}-\alpha \left(\frac{TB}{M-1}\right)\gamma^{\mathrm{maxloss}}_k}.
\end{equation*}

\def \BigInequality{
\begin{equation}
\label{eqBigInequality}
\left(\gamma^\mathrm{maxloss}_k \frac{f_k^{\prime\prime}\left(\alpha \right)f_k\left(\alpha \right)-2{\left(f_k^{\prime}\left(\alpha \right)\right)}^2}{f_k^3\left(\alpha \right)}\right)\left(1-\frac{\gamma^\mathrm{maxloss}_k}{f_k\left(\alpha \right)}\right)
-{\left(\gamma^\mathrm{maxloss}_k\frac{f_k^{\prime}\left(\alpha \right)}{f_k^2\left(\alpha \right)}\right)}^2 < 2\gamma^\mathrm{maxloss}_k\frac{f_k^{\prime}\left(\alpha \right)}{f_k^2\left(\alpha \right)}\left(1-\frac{\gamma^\mathrm{maxloss}_k}{f_k\left(\alpha \right)}\right)
\end{equation}
}
\ifdouble{
\begin{figure*}[!t]
\normalsize
\setcounter{MYtempeqncnt}{\value{equation}}
\setcounter{equation}{47}
\BigInequality
\setcounter{equation}{\value{MYtempeqncnt}}
\hrulefill
\vspace*{4pt}
\end{figure*}
\fi


\section{Proof for Lemma \myref[lem:Convexity2]: Convexity II}
\label{app.convexity.2}
The set in which $\alpha$ is allowed to change, that is $0 \le \alpha \le 1$ is a convex set.
In addition, the minimum of a set of concave functions is concave.
So, what remains is to prove that $r_{w,k},~k\in\boldsymbol{\mathcal{K}}$ are all concave in $\alpha$.

The second derivative of the WIT data rate for the $k$-$th$ WD can be written as
\begin{equation*}
\frac{d^2r_{w,k}}{d{\alpha }^2}=\left(1-\alpha \right)r_k^{\prime\prime}\left(\alpha \right)-2r_k^{\prime}\left(\alpha \right),
\end{equation*}
where primes and double primes are used to represent the first and second derivatives respectively.
Considering (\myref[eqAlphaConstraintEq]), a set of sufficient conditions for $\frac{d^2r_{w,k}}{d{\alpha }^2}<0$ is
\def \eqConv{
\begin{subequations}
\label{eqConv}
\begin{align}
r_k^{\prime}\left(\alpha \right) &\ge 0, \label{eqconvA}\\
r_k^{\prime\prime}\left(\alpha \right) &< 2r_k^{\prime}\left(\alpha \right). \label{eqconvB}
\end{align}
\end{subequations}
}
\eqConv
Inequality (\myref[eqconvA]) is easily proved and seems obvious as the total UL data rate should not decrease as we increase the feedback phase ratio.
To derive a condition under which inequality (\myref[eqconvB]) holds, let us define $f_k\left(\alpha \right)$ as follows
\begin{align*}
f_k\left(\alpha \right)&={\left(1+\gamma^\mathrm{max}_k\right)}^{\alpha \left(\frac{TB}{M-1}\right)+1}-\alpha \left(\frac{TB}{M-1}\right)\gamma^\mathrm{maxloss}_k.
\end{align*}
Using (\myref[eqDataRateWITThree]), we can express inequality (\myref[eqconvB]) in terms of $f_k(\alpha)$ and its derivatives as (\myref[eqBigInequality]).
\ifdouble
\addtocounter{equation}{1}
\else
\BigInequality
\fi
Assuming $\gamma^\mathrm{maxloss}_k>0$, $f_k(\alpha) \neq 0$, and $f_k^\prime(\alpha) \neq 0$, this inequality may be simplified to
\begin{equation*}
\left(\frac{f_k^{\prime\prime}\left(\alpha \right)f_k\left(\alpha \right)}{{\left(f_k^{\prime}\left(\alpha \right)\right)}^2}-2\frac{f_k\left(\alpha \right)}{f_k^{\prime}\left(\alpha \right)}-2\right)\left(f_k\left(\alpha \right)-\gamma^\mathrm{maxloss}_k\right) < \gamma^\mathrm{maxloss}_k,
\end{equation*}
for which a sufficient condition is
$g_{1,k}(\alpha) g_{2,k}(\alpha) <  0,$
where
\begin{align*}
g_{1,k}(\alpha) &= \frac{f_k^{\prime\prime}\left(\alpha \right)f_k\left(\alpha \right)}{{\left(f_k^{\prime}\left(\alpha \right)\right)}^2}-2\frac{f_k\left(\alpha \right)}{f_k^{\prime}\left(\alpha \right)}-2,\\
g_{2,k}(\alpha) &= f_k\left(\alpha \right)-\gamma^\mathrm{maxloss}_k.
\end{align*}
Now, we show that $g_{1,k}(\alpha) < 0$, $g_{2,k}(\alpha) > 0$.

First, it can be shown that for
\begin{equation*}
2\gamma^\mathrm{max}_k<{\ln \left(1+\gamma^\mathrm{max}_k\right)\ }\left(1+\gamma^\mathrm{max}_k\right)
\end{equation*}
to hold, we must have
$\frac{-2}{1+\gamma^\mathrm{max}_k}>W_0\left(-2e^{-2}\right).$
%
Since $W_0\left(-2e^{-2}\right)<0$,
\begin{equation*}
\frac{-2}{W_0\left(-2e^{-2}\right)}-1 < 4 \le \gamma^\mathrm{max}_k.
\end{equation*}
Therefore, for $\gamma^\mathrm{max}_k\ge 4$, and ${\alpha \left(\frac{TB}{M-1}\right)}>0$,
\begin{equation*}
\gamma^\mathrm{maxloss}_k < \frac{1}{2}{\ln \left(1+\gamma^\mathrm{max}_k\right)\ }{\left(1+\gamma^\mathrm{max}_k\right)}^{\alpha \left(\frac{TB}{M-1}\right)+1}.
\end{equation*}
Under the same condition,
\def \eqInequalityCommon{
\begin{equation}
2\left(\frac{TB}{M-1}\right){\ln \left(1+\gamma^\mathrm{max}_k\right)\ }f_k^{\prime}\left(\alpha \right) > f_k^{\prime\prime}\left(\alpha \right).
\label{eqInequalityCommon}
\end{equation}
}
\eqInequalityCommon
Now, let us assume $0<\left(\frac{TB}{M-1}\right)<\frac{1}{{\ln \left(1+\gamma^\mathrm{max}_k\right)\ }}$.
In this case,
\begin{equation*}
f_k\left(\alpha \right) > \frac{f_k^{\prime}\left(\alpha \right)}{\left(\frac{TB}{M-1}\right){\ln \left(1+\gamma^\mathrm{max}_k\right)\ }},
\end{equation*}
and for $\gamma^\mathrm{max}_k\ge 4$,
\ifdouble
\def \eqInequalityOneA{
\begin{equation}
\begin{split}
f_k^{\prime}\left(\alpha \right) &> \frac{1}{2}{\ln \left(1+\gamma^\mathrm{max}_k\right)\ }{\left(1+\gamma^\mathrm{max}_k\right)}^{\alpha \left(\frac{TB}{M-1}\right)+1}\\
&>\frac{1}{2}{\ln \left(1+\gamma^\mathrm{max}_k\right)\ }f_k\left(\alpha \right).
\label{eqInequalityOneA}
\end{split}
\end{equation}
}
\else
\def \eqInequalityOneA{
\begin{equation}
f_k^{\prime}\left(\alpha \right) > \frac{1}{2}{\ln \left(1+\gamma^\mathrm{max}_k\right)\ }{\left(1+\gamma^\mathrm{max}_k\right)}^{\alpha \left(\frac{TB}{M-1}\right)+1}>\frac{1}{2}{\ln \left(1+\gamma^\mathrm{max}_k\right)\ }f_k\left(\alpha \right).
\label{eqInequalityOneA}
\end{equation}
}
\fi
\eqInequalityOneA
On the other hand,
\ifdouble
\def \eqInequalityOneB{
\begin{equation}
\begin{split}
f_k\left(\alpha \right) &> \frac{\ln \left(1+\gamma^\mathrm{max}_k\right){\left(1+\gamma^\mathrm{max}_k\right)}^{\alpha \left(\frac{TB}{M-1}\right)+1}-\alpha \gamma^\mathrm{maxloss}_k}{\ln \left(1+\gamma^\mathrm{max}_k\right)}\\
&> \frac{f_k^\prime(\alpha)}{\left(\frac{TB}{M-1}\right)\ln \left(1+\gamma^\mathrm{max}_k\right)}\\
&> f_k^\prime(\alpha).
\label{eqInequalityOneB}
\end{split}
\end{equation}
}
\else
\def \eqInequalityOneB{
\begin{equation}
f_k\left(\alpha \right) > \frac{\ln \left(1+\gamma^\mathrm{max}_k\right){\left(1+\gamma^\mathrm{max}_k\right)}^{\alpha \left(\frac{TB}{M-1}\right)+1}-\alpha \gamma^\mathrm{maxloss}_k}{\ln \left(1+\gamma^\mathrm{max}_k\right)}
> \frac{f_k^\prime(\alpha)}{\left(\frac{TB}{M-1}\right)\ln \left(1+\gamma^\mathrm{max}_k\right)}\\
> f_k^\prime(\alpha).
\label{eqInequalityOneB}
\end{equation}
}
\fi
\eqInequalityOneB
So, using (\myref[eqInequalityCommon]), (\myref[eqInequalityOneA]), and (\myref[eqInequalityOneB]), as well as the assumption we made, and sensitivity of $\gamma^\mathrm{max}_k$ we can write
$\frac{f_k^{\prime\prime}\left(\alpha \right)}{f_k^{\prime}\left(\alpha \right)}<2,$ and
$1<\frac{f_k\left(\alpha \right)}{f_k^{\prime}\left(\alpha \right)}<2.$
As a result,
\begin{equation*}
g_{k,1}(\alpha) < 0.
\end{equation*}
Now, let us assume $\left(\frac{TB}{M-1}\right)>\frac{1}{{\ln \left(1+\gamma^\mathrm{max}_k\right)\ }}$. In this case
\ifdouble
\def \eqInequalityTwo{
\begin{equation}
\begin{split}
f_k\left(\alpha \right) &< \frac{{\ln \left(1+\gamma^\mathrm{max}_k\right)\ }{\left(1+\gamma^\mathrm{max}_k\right)}^{\alpha \left(\frac{TB}{M-1}\right)+1}-\alpha \gamma^\mathrm{maxloss}_k}{{\ln \left(1+\gamma^\mathrm{max}_k\right)\ }}\\
&< \frac{\alpha f_k^{\prime}\left(\alpha \right)}{\left(\frac{TB}{M-1}\right){\ln \left(1+\gamma^\mathrm{max}_k\right)\ }}.
\label{eqInequalityTwo}
\end{split}
\end{equation}
}
\else
\def \eqInequalityTwo{
\begin{equation}
f_k\left(\alpha \right) < \frac{{\ln \left(1+\gamma^\mathrm{max}_k\right)\ }{\left(1+\gamma^\mathrm{max}_k\right)}^{\alpha \left(\frac{TB}{M-1}\right)+1}-\alpha \gamma^\mathrm{maxloss}_k}{{\ln \left(1+\gamma^\mathrm{max}_k\right)\ }}\\
< \frac{\alpha f_k^{\prime}\left(\alpha \right)}{\left(\frac{TB}{M-1}\right){\ln \left(1+\gamma^\mathrm{max}_k\right)\ }}.
\label{eqInequalityTwo}
\end{equation}
}
\fi
\eqInequalityTwo
So, using (\myref[eqInequalityTwo]), and the facts that both $f_k(\alpha)$ and its derivative are always positive, we can write
\begin{equation*}
0<\frac{f_k\left(\alpha \right)}{f_k^{\prime}\left(\alpha \right)}<\frac{\alpha }{\left(\frac{TB}{M-1}\right){\ln \left(1+\gamma^\mathrm{max}_k\right)\ }}.
\end{equation*}
Using these inequalities and (\myref[eqInequalityCommon]), we can conclude that
$g_{k,1}(\alpha) < 2\alpha -2,$
which means $g_{k,1}(\alpha)$ is always negative.

Now, what remains is to show that $g_{k,2}(\alpha) > 0$.
Assuming $\phi = 1+\alpha \left(\frac{TB}{M-1}\right)>1$, $g_{k,2}(\alpha)$ can be written as
\begin{equation*}
g_{2,k}(\phi) = {\left(1+\gamma^\mathrm{max}_k\right)}^\phi-\beta\gamma^\mathrm{maxloss}_k.
\end{equation*}
Using its first and second derivative, it can be shown that the minimum of this function is $g_{2,k}(1) = 1+\gamma^\mathrm{max}_k-\gamma^\mathrm{maxloss}_k$ which is positive.



\ifextended
\newpage

\color{olive}

\section{Notes}

\subsection{Reviews Items}
\begin{itemize}
\item How do we show the elements of a vector? This is used in 33 and 34
\item Check the recommended elsevier paper in Globecom
\item distinguishing the paper from its predecessors especially the IET paper
\item Appendix C and D seem to be placed in the wring order!
\item Check whether or not an Equation should be in math mode
\item Check if Equation should be numbered or not 
\item Check active vs. Passive voice of all sentences
\item Convert some calars to capitals
\item Check Vocabulary
\item We will be using vector and scalar notations alternatively
\item Citation correctness
\item Checking equation references
\item Consistency of variable names
\item Low-level math
\item High-Level math
\item Paragraph thesis statement
\item Important author Citation in 1st page
\item Intro, abstract, Conclusion
\item Proof Endings
\item Order of citations
\item Re-run simulations for 1000 runs
\item MIMO emphasis
\item Checking Upload and Download Channel Matrices
\item Check "given by" "written as" etc.
\item Introduce elements of vectors everywhere!
\item Using $b_u$ and $b_d$
\item Using efficiency
\item Column vectors assumed
\item Transform some sentences to footnotes
\item Check "the" before publication
\item Grammar
\item Subject order
\item MU-MISO trade-off in feedback in addition to the traditional WPT, WIT trade-off
\item Long equations could be placed in a single-column format
\end{itemize}
-----------------------------------------------------------------------
\subsection{Labels}
\subsubsection{Figures}
\begin{itemize}
\item fig.schematic
\item fig.frame
\item fig.sim1
\item fig.sim2
\item fig.sim3
\end{itemize}
-----------------------------------------------------------------------

\subsubsection{Appendices}
\begin{itemize}
\item app.UL.data.rate
\item app:asymptotic.alpha
\item app.convexity.1
\item app.convexity.2
\item app.pareto.optimal.beamformer
\end{itemize}
-----------------------------------------------------------------------

\subsubsection{Corollaries}
\begin{itemize}
\item cor.asymptotic.alpha
\item cor.asymptotic.beta
\item cor.asymptotic.r.f
\item cor.asymptotic.xiup
\end{itemize}
-----------------------------------------------------------------------

\subsubsection{Lemmas}
\begin{itemize}
\item
\end{itemize}
-----------------------------------------------------------------------

\subsubsection{Theorems}
\begin{itemize}
\item
\end{itemize}
-----------------------------------------------------------------------

\subsection{Equations}
\begin{itemize}
\setcounter{equation}{0}

\item eqLongTermFading
\eqLongTermFading

\item eqULSignal
\eqULSignal

\item eqZFDetector
\eqZFDetector

\item eqDetectedSignalVector
\eqDetectedSignalVector

\item eqDetectedSignalScalar
\eqDetectedSignalScalar

\item eqSINROne
\eqSINROne

\item eqShannon
\eqShannon

\item eqSINRTwo
\eqSINRTwo

\item eqSINRThree
\eqSINRThree

\item eqDataRateWIT
\eqDataRateWIT

\item eqFeedbackErrorOne
\eqFeedbackErrorOne

\item eqFeedbackErrorTwo
\eqFeedbackErrorTwo

\item eqDLSignal
\eqDLSignal

\item eqEnergyVectorOne
\eqEnergyVectorOne

\item prParetoOptimalBeamformer
\prParetoOptimalBeamformer

\item eqParetoOptimalBeamformer
\eqParetoOptimalBeamformer

\item eqEnergyVectorTwo
\eqEnergyVectorTwo

\item eqChannelOuterProd
\eqChannelOuterProd

\item eqEnergyScalar
\eqEnergyScalar

\item eqLambdaOne
\eqLambdaOne

\item eqMDecomposition
\eqMDecomposition

\item eqLambdaMax
\eqLambdaMax

\item eqLambdaLossOne
\eqLambdaLossOne

\item eqLambdaLossTwo
\eqLambdaLossTwo

\item eqLambdaMaxloss
\eqLambdaMaxloss

\item eqFundamentalOne
\eqFundamentalOne

\item eqFeedbackErrorThree
\eqFeedbackErrorThree

\item eqDataRateWITThree
\eqDataRateWITThree

\item prOptimizationOne
\prOptimizationOne

\item eqDataRateWITTwo
\eqDataRateWITTwo

\item eqMDiag
\eqMDiag



\item eqSINREquality
\eqSINREquality

\item eqKSet
\eqKSet

\item eqOptimalXiupOne
\eqOptimalXiupOne

\item eqSINREquality
\eqSINREquality

\item eqRMaxKSet
\eqRMaxKSet

\item eqFairnessRadiusOne
\eqFairnessRadiusOne


\arx{
\item eqFairnessRadiusTwo
\eqFairnessRadiusTwo
}

\item eqOptimizationTwo
\eqOptimizationTwo

\item eqOptimalBeta
\eqOptimalBeta

\item eqOptimalAlphaOne
\eqOptimalAlphaOne

\item eqOptimalAlphaTwo
\eqOptimalAlphaTwo

\item eqAsymptoticXiup
\eqAsymptoticXiup

\item eqAsymptoticFairnessRadius
\eqAsymptoticFairnessRadius

\item eqAsymptoticBeta
\eqAsymptoticBeta

\item eqAsymptoticAlpha
\eqAsymptoticAlpha

\item eqConv
\eqConv

\item BigInequality
\BigInequality

\item eqInequalityCommon
\eqInequalityCommon

\item eqInequalityOneA
\eqInequalityOneA

\item eqInequalityOneB
\eqInequalityOneB

\item eqInequalityTwo
\eqInequalityTwo

\end{itemize}

\fi

\end{document}